\documentclass[aps,preprint,showpacs ,amssymb, amsmath,pre]{revtex4-1}
\usepackage{bm}
\usepackage{graphicx}

\newcommand{\bdm}{\begin{displaymath}}
\newcommand{\edm}{\end{displaymath}}
\newcommand{\beq}{\begin{equation}}
\newcommand{\eeq}{\end{equation}}
\newcommand{\beqa}{\begin{eqnarray}}
\newcommand{\eeqa}{\end{eqnarray}}

\newcommand{\edo}{\end{document}}

\newcommand{\bit}{\begin{itemize}}
\newcommand{\eit}{\end{itemize}}
\newcommand{\ben}{\begin{enumerate}}
\newcommand{\een}{\end{enumerate}}

\newcommand{\rar}{\rightarrow}

\newcommand{\bfP}{{\bf P}}
\newcommand{\bfp}{{\bf p}}
\newcommand{\bfA}{{\bf A}}
\newcommand{\bfK}{{\bf K}}

\newcommand{\bfbe}{\bm{\beta}}

\newcommand{\bfa}{{\bf a}}
\newcommand{\bfd}{{\bf d}}
\newcommand{\bfn}{{\bf n}}

\newcommand{\bfr}{{\bf r}}

\newcommand{\bfk}{{\bf k}}

\newcommand{\bfv}{{\bf v}}

\newcommand{\viq}{\, , \qquad}

\begin{document}

\title{Thomson and Compton scattering with  an intense laser pulse}

\author{Madalina Boca}
\email{madalina.boca@g.unibuc.ro}
\affiliation{Department of Physics  and Centre for Advanced Quantum Physics,  University of Bucharest, MG-11, Bucharest-M\u agurele, 077125  Romania}

\author{Viorica Florescu }
\affiliation{Department of Physics  and Centre for Advanced Quantum Physics,  University of Bucharest, MG-11, Bucharest-M\u agurele, 077125  Romania}

\begin{abstract}

Our paper concerns the scattering of intense laser radiation on free  electrons and it is  focused on the relation between nonlinear Compton and nonlinear Thomson scattering. The analysis is performed for a laser field modeled  by an ideal pulse with a finite duration, a fixed direction of propagation and indefinitely extended in the plane perpendicular to it. We  derive the classical limit of  the quantum  spectral and angular distribution of the emitted  radiation,    for an arbitrary polarization of the  laser  pulse.   We also rederive our result directly, in the framework of classical electrodynamics,   obtaining, at the same time, the distribution for the emitted  radiation with a well defined  polarization.   The  results  reduce to those established by   Krafft {\it et al.}  [G. A. Krafft, A. Doyuran and J. B. Rosenzweig,  Phys. Rev. E {\bf 72}, 056502 (2005)] in the particular  case of linear polarization of the pulse, orthogonal to the initial electron momentum. Conditions in which  the differences between classical and quantum results  are visible are discussed and illustrated  by graphs.

\end{abstract}

\maketitle

\section{Introduction}\label{s-i}

The invention of the laser fifty years ago, has fostered many theoretical studies  \cite{Eberly}, treating the interaction of the electrons with an {\it intense electromagnetic field}. For theorists this was a field   where new methods of investigation were necessary.  Based on these,  processes such as electron reflection and refraction, nonlinear Compton scattering and one-photon  pair production have been studied. In recent years, progress of experimental physics,  leading to the detection of some of the fundamental nonlinear processes possible  in head-on collisions of very fast electrons with an intense laser beam (10$^{18}$ W/cm$^2$) \cite{SLAC}, has renewed interest in  theoretical studies, especially in the relativistic regime. Review papers such as those of Salamin {\it et al.} \cite{Hatz} and Ehlotzky {\it et al.} \cite{EKK}  have been published and we refer to them for bibliography and details.

The particular new interest for  {\it nonlinear Compton scattering} with free electrons is also related to the possibility it opens for new sources of ultra short pulses  in the X-ray domain, with durations extending from  the picosecond range, as discussed by Esarey {\it et al.} \cite{Esarey}, to the attosecond domain  \cite{as}.  Even the production of zeptosecond X-ray pulses is under theoretical study \cite{LLCW}. Envisaged applications are numerous and various  and they  continue to stimulate experimental investigations \cite{Bab}.

In the  description of intense {\it radiation scattering}  on free electrons,    {\it classical electrodynamics} (CED), in which the electron  is subject to the laws of classical mechanics and the electromagnetic field obeys Maxwell equations, as well the  theory applying  quantum mechanics for the electron, are  used in the literature. 
 In the latter case, the external field is described classically but,  as the  process involves spontaneous emission of  a photon,  the interaction with the quantized  electromagnetic field is also taken in account. This hybrid approach will be referred to as  {\it quantum}  in the following. 

In the regime of low  incident radiation frequency $\,\omega_1\,$,  such that $\, \hbar \omega_1 \ll  mc^2\,$ ($m$ the electron mass and $c$ the velocity of light), the name {\it Thomson scattering} is currently given to the process. In this situation, classical theory is  used and so under the name of nonlinear Thomson scattering one finds classical calculations. And then one speaks sometimes in the literature about "the transition from Thomson to Compton scattering" \cite{MKM}. We shall adhere to this terminology in our paper, using Compton's  name  when  referring to  quantum calculations and Thomson's name for classical calculations.

The majority of published calculations, classical or quantum, refer to the {\it monochromatic case}.    With some exceptions that   illuminate the  analysis of the laser pulse case,    we shall not quote them here, as they are well described  in the review papers we have mentioned before.

 The subject of our paper is {\it the  connection} between Thomson and Compton scattering {\it beyond the monochromatic case}.  Our study is performed for  an  electromagnetic plane wave which is supposed to have a finite extension in the direction of propagation but an indefinite extension in the plane orthogonal to this direction. This wave will be designated in the following as a {\it laser pulse}. Our contribution concerns two aspects: i) the connection between the  basic equations describing Thomson and Compton   scattering and ii) qualitative and quantitative similarities and  differences in the predictions extracted from these equations.

For {\it Thomson scattering with a laser pulse}, alternative expressions to  the  general ones found in the textbooks \cite{Jks} for the energy and angular distribution of the emitted radiation have been derived by Krafft \cite{Krf1, Krf2}  in a particular scattering geometry. The evolution of the classical electron in the case of a plane wave and beyond it is discussed in detail by Hartemann and coworkers \cite{Hart3D} in their  paper presenting calculations  based on an accurate description of the three-dimensional focus of
a laser wave, in both the near-field and far-field regions;  more  papers presenting  numerical  calculations for a realistic, focused laser pulse  have { been published in the last ten years (Lee {\it et al.}  \cite{cor}, Gao \cite{Gao}, Lan {\it et al.} \cite{Lan}, Heinzl et al. \cite{UK-nou}).  Other  numerical calculations \cite{Hartpro} have investigated different aspects such as    the effects of the electron beam emittance and energy spread on the emitted spectrum. More subtle effects as radiation damping are studied \cite{HG}. 

Although elaborated many years ago, {\it quantum theory} have  been applied less to this problem  and  almost all the calculations refer only to an electromagnetic plane wave,  taking advantage of the existence of  the Volkov solutions of the Dirac equation. Several recent extended calculations  have been made for the    monochromatic  case \cite{UK, HI} and some others  for  the laser pulse case \cite {Fofa,BF,UK-nou}.   A quantum calculation  that goes beyond the electromagnetic plane  wave model for the laser beam and the description by Volkov solutions is that of 
Krekora {\it et al.} \cite{Krekora}. They  have solved the time-dependent Dirac equation for an electron wave packet accelerated by a
very strong laser field and have  used the Lienard-Wiechert solution of  Maxwell equations in order to calculate the scattered
light spectra. In particular, they have shown that the width of the  initial  electronic wavepacket influences the spectrum. More recently,  the emission by an electron described by a wavepacket was studied by  Peatross {\it et al.} \cite{wpk}, with the conclusion that "the radiative response can be
mimicked by the incoherent emission of a classical ensemble
of point charges".

 Generally, it is expected that the CED results are good when the electron energy loss is small, such that its effect on the electron motion is negligible. This situation was already exploited decades  ago by Kramers \cite{Kramers} in a study of electron  bremsstrahlung in a Coulomb field and  leads to results close to those given by quantum theory \cite{RHP}  for small kinetic energy $\,T_1\,$ of the electron, $\,T_1\ll \alpha^2\,Z^2\,,$  with $\,\alpha\,$ the fine structure constant and $\,Z\,$ the nuclear charge.

The connection between quantum and CED results has been discussed in the literature in the monochromatic case. In particular, Goreslavskii {\it et al.} \cite{GRSL} have derived the classical energy and angular distribution of radiation starting  from the quantum results, in the case  of circularly polarized monochromatic radiation. They formulate conditions of applicability of the classical limit  for weak fields  and for very intense fields. Very  recent papers comparing  Thomson and Compton scattering are those of Hartemann \cite{Hart} and Heinzl {\it et al.} \cite{UK-nou}.  In \cite{UK-nou} the comparison is  made using  numerical data obtained from both classical and quantum calculations   for the  particular conditions relevant to  experiments
planned at the Forschungszentrum Dresden-Rossendorf (FZD) and a set of  realistic parameters is proposed for an experiment allowing the detection of non-linear Thomson scattering to be performed.

New features of Compton scattering at intensities of the order 10$^{18}$W/cm$^2$ and very energetic electrons (GeV range) have been revealed by  recent studies \cite{Heinzl}.  At intensities above 10$^{22}$W/cm$^2$ predictions are that QED effects will modify the radiation emission from these electrons \cite{Sokol}.

Our paper is organized as follows.  At the beginning of Sect. \ref{s-ii} we describe the theoretical framework in which  our study is performed  and the  equations we use in  several  analytic  calculations in both quantum (Sect. \ref{s-ii} A) and classical (Sect. \ref{s-ii} B) treatments. In Sect. \ref{s-iii}  we present, for the case of a laser pulse,  the derivation of an alternative expression (\ref{class}) to the standard  one [Eq. (14.65) of Jackson's book \cite{Jks} ]  for the energy and angular distribution of the emitted  radiation.  This expression  is derived as the classical limit of the quantum expression (\ref{quantum}). Based on the calculation in Appendix \ref{sa-b}, made in the framework of CED, we present the classical radiation  distributions for two particular state of polarizations. Our formulas are valid for any polarization of the laser;  for the particular case of linear polarization orthogonal to the direction of the incident electron they reduce  to the  expressions  published by Krafft {\it et al.} \cite{Krf2} which are valid only  for this configuration.    Sections  \ref{s-iv} and \ref{s-v} are devoted to the comparison between classical and quantum calculations.  While  qualitative aspects are discussed in Sect. \ref{s-iv} implying also the monochromatic case, in Sect. \ref{s-v} the comparison is quantitative and it is made with the aim of revealing situations in which  the differences matter.

The formulae are given in S.I. units. Numerical results for photon distributions are presented in atomic  units,  but photon energies are given in eV.  The equations in Appendix \ref{sa-a} for the classical electron  trajectory  are used in Sect. \ref{s-ii} in the passage from quantum to classical distributions and in Appendix \ref{sa-b}, where the  new version of the classical distribution is derived directly from Eq. (14.65) of Jackson's textbook \cite{Jks}.

\section{Theoretical background}\label{s-ii}

Compton scattering is the name primarily associated with the interaction of a free electron and a photon leading to the emission of a photon with a different direction and frequency (inelastic scattering) and the recoil of the electron. Energy and momentum are conserved and this leaves undetermined two of the six variables of the problem (final photon and electron momenta). This picture, implying the photon,  is justified  in QED, using second-order perturbation theory,  and is suitable at low intensities of the external field. Quantum theory also describes the case in which the electromagnetic field is intense, when perturbation theory is not applicable. In this case the theory  adopts  a hybrid approach, describing the electron within quantum mechanics (Dirac equation),  the external  electromagnetic  field within CED, and also takes into consideration the interaction of the electron with the quantized electromagnetic field which is responsible for spontaneous photon emission.  One can single out the rate for the emission of only one photon in the interaction of the electron   with the two electromagnetic fields.

In the following the external field is a plane wave with a fixed direction of propagation characterized by a unity vector  $\,\bfn_1\,,$ with a finite extension in the direction of propagation and indefinitely extended in the transverse plane. The initial electron momentum is denoted by $\,\bfp_1\,$. The final electron has a momentum denoted by $\,\bfp_2\,$ and the emitted photon has the frequency $\,\omega_2\,$ and the  propagation direction  along the unity vector $\,\bfn_2\,$.

A  basic quantity predicted by    quantum  theory is  the   radiated energy  distribution over  frequency and direction of the emitted photon and over the momentum of the scattered electron
\beq
d^4\,W= \rho_4(\omega_2,\bfn_2,\bfp_2)\,d\omega_2\,d\Omega_2\,d\bfp_2\,.
\eeq
We refer to the case in which an average and a summation over the  initial  and final electron spin was made, and a summation over the  emitted photon polarization as well.  If the scattered electron is not observed, the relevant quantity is
\beq
d^2\,W=\rho_2(\omega_2,\bfn_2)\,d\omega_2\,d\Omega_2\viq \rho_2(\omega_2,\bfn_2)=\int_{\bfp_2}
  \rho_4(\omega_2,\bfn_2,\bfp_2)\,d\bfp_2\,.
\eeq

In CED formalism, the electron, accelerated by the field, emits radiation. The theory leads to predictions for this radiation. The calculation we refer to is made with the assumption that the emission of radiation does not affect the particle motion, so the  electron moves on a trajectory well determined by  the initial conditions and  the final electron momentum is well determined.  As a consequence one is led only to a distribution of the emitted radiation frequency and direction,
\beq
d^2\,W^{\mathrm{cl}}=\rho_2^{\mathrm{cl}}(\omega_2,\bfn_2)\,d\omega_2\,d\Omega_2\,.
\eeq

From the distribution $\,d^2W/d\omega_2\,d\Omega_2\,$ one obtains,  in either the classical or the quantum case,  the angular distribution of the radiated energy $\,dW/d\Omega_2\,$, its  frequency distribution $\,dW/d\omega_2\,$ and the total emitted energy $\,W_{\mathrm{total}}\,$.

The external field we consider is described by a vector potential  $\,{\bf A}\,$ orthogonal to the direction of propagation ${\bf n}_1$ and depending only on the variable
\beq\label{var}
\phi=t-\bfr\cdot\bfn_1/c\,.
\eeq
We shall use the  four-vectors
\beq\label{n1n}
n_1=(1,\bfn_1)\viq \widetilde n_1=(1,-\bfn_1)\viq n_1^2=\widetilde n_1^2=0\viq n_1\cdot\widetilde n_1=2\,,
\eeq
so  $\,\phi=n_1\cdot x/c\,$, where $\,x =(ct,\bfr)\,$ is the position four-vector.
In the monochromatic case the  vector potential $\,\bfA\,$ is a periodic function of $\,\phi\,$, while in  the {\it laser pulse} case it could be an arbitrary function of it, which 
 becomes negligible for values of $\,\phi\,$  outside a finite range $\,(\phi_{\mathrm{in}}, \phi_{\mathrm f})\,$. This implies that  at fixed position $\,\bfr\,$  the external field is present a finite duration, equal to $\,\phi_{\mathrm f}-\phi_{\mathrm{in}}$. We denote by $\,A_0\,$ the maximum value taken by the vector potential and use the notation 
\beq
\eta\equiv\frac{\mid e\mid A_0}{mc}\label{defeta}\,,
\eeq
with $\,e<0\,$ the electron charge. In the monochromatic case the 
 dimensionless quantity $\,\eta^2\,$     is directly proportional to the intensity of the wave and inversely proportional  to the square of the frequency.

\subsection{The radiated energy distribution in quantum theory}

As mentioned before, in the   approach   we use   here,    the external field is described classically; the electron, described in quantum mechanics, interacts also with the quantized electromagnetic field and this interaction,  treated in first-order perturbation theory, leads to spontaneous emission of radiation. 

 Owing to the existence of the Volkov spinors,  exact solutions of the Dirac equations for the electron in a classical  electromagnetic plane wave, which reduce to plane wave spinors in the absence of the field,  one can write an analytic expression for the  {\it transition amplitude}. Its expression  in the monochromatic case is given in many papers, starting with those from 1964, as quoted for instance in \cite{EKK}.  For the case of a pulse, the transition amplitude is proportional  to  a product of $\,\delta$-functions
\beq\label{amp}
{\cal T}\sim\,\delta({\bf p}_{1\perp}-{\bf p}_{2\perp}-\hbar {\bf k}_{2\perp})\,\delta[ n_1\cdot(p_1-p_2-\hbar k_2)]\,,
\eeq  
where the index $\,\perp\,$   indicates the components orthogonal to the direction of propagation of the laser pulse ${\bf n}_1$, taken in the following as the $Oz$ axis.
The function that multiplies the  $\,\delta$-functions can be expressed in terms of three integrals  \cite{BF}. Details will given in what follows.

As a consequence of the $\,\delta$-functions present in (\ref{amp}), three conservation laws are valid in the case of a laser pulse. They completely determine the value of the momentum $\,\bfp_2\,$, for fixed emitted photon frequency and direction. According to Eq. (40) of \cite{BF}, this momentum, denoted by $\,\widetilde \bfp_2\,$, has the spatial components
\begin{equation}
\widetilde{\bf p}_{2\perp}={\bf p}_{1\perp}-\hbar{\bf k}_{2\perp},\qquad \widetilde{ p}_{2z}=\frac{(mc)^2+({\bf p}_{1\perp}-
\hbar{\bf k}_{2\perp})^2}{2n_1\cdot (p_1-\hbar k_2)}-\frac{n_1\cdot(p_1-
\hbar k_2)}2\,\label{consp}\,
\end{equation}
and the temporal component  
\beq  
\widetilde p_2^0=\sqrt{(mc)^2+ \widetilde \bfp^2}=\frac{(mc)^2+({\bf p}_{1\perp}-\hbar{\bf k}_{2\perp})^2}{2n_1\cdot(p_1-\hbar k_2)}+\frac{n_1\cdot(p_1-\hbar k_2)}{2}\,.\label{cons3}
\eeq
 The conservation law for the momentum components perpendicular to $\,{\bf n}_1\,$ is the same as  in the   monochromatic case.

The three 1D $\delta$-functions are also present in the transition probability. 
Integration over the electron attributes eliminates  them. 
 In order to obtain the probability for the emission of a photon with the frequency in $(\omega_2,\omega_2+d\omega_2)\,$ and direction in  the solid angle element $\,d\Omega_2\,$, 
we can directly use Eqs. (42) in \cite{BF}: we have to multiply the cross section $\,d^2\sigma_\gamma\,$   by  the flux $\,J\,$ (as defined in Eq. (33)  of the quoted paper)  and  by  the effective duration of the pulse  $\,\widetilde \tau_p\,$.   We also have to include  the factor $\,\widetilde p_2^{\,0}/n_1\cdot\widetilde  p_2\,$  omitted previously in \cite{BF}. After that, the quantum radiated energy distribution is obtained  by    multiplying the transition probability with the energy of the  emitted photon.  

Here we write the spectrum in terms of the integrals $\,{\bf a}^{(q)}\,$ (already used in \cite{BF}, and denoted there by $\,{\mbox{\boldmath${\cal{A}}$}}\,$),  and the integral $\,{ b}^{(q)}\,$ we introduce here. These integrals are 
\beqa\label{ABq}
{\bf a}^{(q)}&=&-\int\limits_{-\infty}^{\infty}\,d\phi\,\frac{e{\bf A}(\phi)}{mc}\,\exp\lbrack\,-{ i}
\,\Phi(p_2,p_1;\phi)\,\rbrack\,,\;\quad\label{integrA}\\
 {b}^{(q)}&=&
\int_{{-\infty}}^{\infty}\,d\phi\,\frac{e^2\bfA^2(\phi)}{2(mc)^2}\exp\lbrack\,
-{ i}\,\Phi(p_2,p_1;\phi)\,\rbrack\,,\label{integrB0}
\eeqa
with
\beq\label{faza}
\Phi(p_2,p_1;\phi)\equiv \frac{1}{\hbar}\,\left[\frac{c}{2}\,\phi\,\widetilde n_1\cdot(p_1-p_2-\hbar k_2)+F(p_1;\phi)-F(p_2;\phi)\right]
\eeq
and
\beq\label{fj}
F(  p;\phi)=\frac c{2( n_1\cdot p)}\int\limits_{\phi_0}^{\phi}d\chi[e^2{\bf A}^2(\chi)-2e{\bf A}(\chi)\cdot{\bf p}_{\perp}]\,.
\eeq
The integration  in (\ref{integrA}) and (\ref{integrB0}) extends in fact over the pulse duration.  A change of the  value of the  arbitrary constant $\,\phi_0\,$ in (\ref{fj}) will lead to an over-all phase factor in the transition amplitude with no effect on the transition probability.

 As a result of the integration over the electron momentum, in the expression of the photon spectral and angular distribution that follows,  the four momentum $\,p_2\,$ has to be  replaced with the four-momentum $\,\widetilde p_2\,$  described in (\ref{consp}). We write the mentioned distribution as
\begin{equation}\label{quantum}
\frac{d^2W}{d\omega_2\,d\Omega_2}=\frac{e_0^2\omega^2}{4\pi^2c}\frac{(mc)^2}{(n_1\cdot \widetilde p_2)^2}\frac{\widetilde p_2^{\,0}}{p_1^{\,0}}{\cal R}\,,
\end{equation}
 where
\begin{equation}
{\cal R}=c_1\,|{\bf a}^{(q)}|^2+{c_2\,(mc)^2\vline\,\frac{{\bf a}^{(q)}\cdot{{\bf d}}}{{\bf n}_1\cdot{\bf P}}\,\vline}^{\,2}+c_3\,|b^{(q)}|^2+2\,c_4\,m\,c\,\Re\left((b^{(q)})^*\,\frac{{\bf a}^{(q)}\cdot{\bf d}}{{\bf n}_1\cdot{\bf P}}\right)\,,\label{defR}
\end{equation}
 with
\begin{eqnarray}
c_1&=&1+\frac{(\hbar k_2\cdot n_1)^2}{2(mc)^2}\gamma_{11}\viq
c_2=2\,{\bf n}_1\cdot{\bf P}\,\frac{c_1}{\gamma_{11}(n_1\cdot \hbar k_2)}+c_5\nonumber\\
c_3&=&(mc)^2\,\frac{ d_0^2}{({\bf n}_1\cdot{\bf P})^2}\,c_5\viq
c_4= d_0\,\lbrack\,\frac{mc}{n_1\cdot \hbar k_2}\frac{c_1}{\gamma_{11}}+\frac{mc}{{\bf n}_1\cdot{\bf P}}\,c_5\,\rbrack\,,\label{c14}
\end{eqnarray}

 and
\begin{eqnarray}
c_5&=&\left[-1+\left(\frac1{\gamma_{12}}+\frac1{\gamma_{21}}\right)\frac1{n_1\cdot n_2}\right],\quad\gamma_{ij}=\frac{(mc)^2}{(n_i\cdot p_1)(n_j\cdot\widetilde p_2)},\quad i,j=1,2\label{defP}\\
{d}_0&=&m\,c\,\left(\frac1{n_1\cdot p_1}-\frac1{n_1\cdot \widetilde p_2}\right),\quad  {\bf d}=\frac{{\bf p}_1}{n_1\cdot p_1}-\frac{\widetilde{\bf p}_2}{n_1\cdot {\widetilde p_2}},\quad  {\bf P}={\bf p_1}-\widetilde{\bf p}_2-\hbar{\bf k}_2 \;.\label{defd}
\end{eqnarray}

\subsection{The classical energy and angular distribution}

The most used equation for the distribution of the emitted photons over directions and frequencies is that derived in Jackson's book [see \cite{Jks},  Eq.(14.65)], 
 \begin{equation}
  \label{eq:d2W}
  \frac{d^2W^{\mathrm {cl}}}{d\omega_2 d\Omega_2}=2\, | {\bf K}(\omega_2,\bfn_2)|^2\,,
\end{equation}
where
\begin{equation}
  \label{eq:S}
 {\bf K}(\omega_2,\bfn_2)=\frac{e_0}{2\pi\sqrt{2 c}}\int\limits_{-\infty}^{\infty}\,\frac{dt}{\kappa^2}\,{\bf n}_2\times[({\bf n}_2-\bfbe)\times\dot{\bfbe}]\,\exp[-i\Phi^{\mathrm {cl}}]\viq e_0^2\equiv\frac{e^2}{4\pi\epsilon_0}\,,
\eeq
with $\,{\bf n}_2$  the unity vector of the observation direction and 
\begin{equation}
  \label{fazacl}
\bfbe(t)=\frac{\bfv(t)}{c}\viq   \kappa=1-{\bf n}_2\cdot \bfbe\viq \Phi^{\mathrm {cl}}=-\omega_2\,\left(t-\frac{{\bf n}_2\cdot{\bf r}(t)}{c}\right)\,.
\end{equation}
In  formula (\ref{eq:S})   the position $\,\bfr(t)\,$, the   velocity   and the acceleration of the moving electron are all implied.

If the {\it electron is at rest at the initial and final moments} in regions where the  electromagnetic pulse is zero, an integration by parts  leads to the simpler expression [Jackson's Eq. (14.67)],
\begin{equation}
  \label{eq:S-red}
  {\bf K}(\omega_2,\bfn_2)=-{ i}\,\frac{e_0}{2\pi\sqrt{2 c}}\,\omega_2\,\int\limits_{-\infty}^{\infty}\,dt\;{\bf n}_2\times({\bf n}_2\times \bfbe)\,\exp[-i\Phi^{\mathrm{cl}}]\,,
\end{equation}
which requires  only the electron position and velocity along its trajectory.

 Equations (\ref{eq:S}) and (\ref{eq:S-red}) are valid for any external electromagnetic field.  In the following we shall refer only to {\it the case of a plane wave}, when the only variable is $\,\phi\,$, defined  in (\ref{var}). In this case the  expressions given in Appendix \ref{sa-a} for the trajectory and velocity are valid.   We transcribe  here   the expression  (\ref{fazacl}) of the classical phase  $\,\Phi^{\mathrm{cl}}\,$, as convenient for further reference. First we write
\beq
 \Phi^{\mathrm{cl}}=-\omega_2\,\lbrack\, \phi+(\bfn_1-\bfn_2)\cdot\frac{\bfr(t)}{c}\,\rbrack\,
\eeq
and then, using the expressions of the vector $\,\bfr\,$,  as a function of $\phi\,$ as in Eqs.(\ref{traj}),  one obtains the phase in terms of  the variable  $\,\phi\,$. Starting from now, we shall  use the notation $\,\Phi^{\mathrm{cl}}(p_1;\phi)\,$ displaying also the argument $\,p_1\,$, but not $\,\omega_2\,$ and $\,n_2\,$, in order to have a reasonably simple notation, corresponding to that used in (\ref{faza}), 
\beq\label{fazacltr1}
\Phi^{\mathrm{cl}}(p_1;\phi)=-\omega_2\,[\,\phi\,\frac{n_2\cdot p_1}{n_1\cdot p_1}+\frac{n_1\cdot n_2}{2(n_1\cdot p_1)^2}\,\int_{\phi_{0}}^\phi e^2\bfA^2(\chi) \,d\chi+{\bf W}\cdot \int_{\phi_{ 0}}
^\phi \frac{e\bfA(\chi)}{mc}\,d\chi]+\Phi_0\,,
\eeq
with 
\beq
{\bf W}\equiv mc\,\left(\frac{\bfn_2}{n_1\cdot p_1}-\frac{n_1\cdot n_2}{(n_1\cdot p_1)^2}\,\bfp_1\right)\,.\label{W}
\eeq
$\,\Phi_0\,$  is a constant with respect to the integration variable $\,t\,$ in  (\ref{eq:S}).  It gives a constant phase factor  which  has no effect on the energy distribution.

A change of variable from $\,t\,$ to $\,\phi\,$ is useful and  is described in Appendix \ref{sa-b}. The variable $\,\phi\,$ has already been used in the literature  both in quantum and classical calculations \cite{BK,Sa-Sa}; it was used  in particular by  Krafft {\it et al.} \cite{Krf1,Krf2} in their derivation of the energy distribution for two polarization states of the emitted photon, for a special initial configuration.   In the next section, using  a procedure  different from that   Krafft {\it et al.},  we derive the same type of formula, but  valid for any polarization of the laser and arbitrary initial directions of propagation of the laser and initial electron.

In CED, a general compact expression is valid for the angular distribution (Eq. (14.53) of \cite{Jks}), namely
\beq\label{ang}
\frac{dW^{\mathrm{cl}}}{d\Omega_2}=\frac{e_0^2}{4\pi c}\int\limits_{-\infty}^{\infty}\,\frac{dt}{\kappa^5}\,|\,{\bf n}_2\times[({\bf n}_2-{\bm{\beta}})\times\dot{\bm{\beta}}]\,|^{\,2}\,,
\eeq
with $\,\kappa\,$ defined in (\ref{fazacl}).  In fact, Jackson's calculation \cite{Jks} starts with the angular distribution of the radiation and from it derives  its   spectral decomposition. This means that going from  Eq. (\ref{eq:d2W}) to Eq. (\ref{ang}), i.e.,  integrating over emitted  frequencies,   the interferences between the fields emitted at different times  are destroyed. We have used Eq. (\ref{ang}) in the verification of our numerical codes.

\section{Alternative expression for the  energy distribution of radiation emitted in Thomson scattering  with a laser pulse}\label{s-iii}

We present here how the classical energy distribution emerges from the quantum expression  (\ref{quantum}).  That our equation (\ref{class}) can be obtained also from the general formulas of CED is proven in Appendix \ref{sa-b}. At the end of this section we shall  present the generalization of the analytic results derived  previously by Krafft {\it et al.} \cite{Krf2}.

 The classical limit can be obtained in a very formal way as the limit $\,\hbar  \rar 0\,$ of the quantum expression (\ref{quantum}). The Planck constant appears explicitly,  as $\,1/\hbar\,$ in the expression (\ref{faza})  of the phase, and   through the emitted photon momentum  $\,\hbar\,k_2$.  We expect the classical approximation to be good if the emitted photon energy $c\hbar k_2$ is small with respect to the electron energy.

Firstly we analyze the phase (\ref{faza}). We need  to consider the phase for the value $\,\widetilde p_2\,$ of the first argument, i.e.,  $\,\Phi(\widetilde p_2,p_1;\phi)\,$, with $\,\widetilde p_2\,$ given by (\ref{consp}) and (\ref{cons3}). In this case  one has
\beq
\widetilde n_1\cdot (p_1-\widetilde p_2-\hbar\,k_2)=2\,\bfn_1\cdot {\bf P}\,
\eeq 
where $\,\widetilde n_1\,$ and $\,{\bf P}\,$ are  defined in (\ref{n1n}) and (\ref{defP}), respectively.  The general expression (\ref{faza}) of the phase can then be directly transcribed as   
\beq
\Phi(\widetilde p_2,p_1;\phi)=\frac{1}{\hbar}\,\lbrack \,c \phi\,\bfn_1 \cdot \bfP+\frac{e^2}{2m}\, d_0\,\int_{\phi_0}^{\phi}\,d\chi\,\bfA^2(\chi) -ec\, \bfd\,\cdot \int_{\phi_0}^\phi \,d\chi\,\bfA(\chi)\,\rbrack\,,
\eeq
with $\, d_0\,$ and $\,{\bf d}\,$ defined in (\ref{defd}).
We keep only the terms linear in $\,\hbar k_2\,$ in $\, d_0\,$ and $\, \bfd\,$. It is easy to show that
\beqa
c\,\bfn_1\cdot\bfP&=&-\hbar\,\omega_2\,\frac{n_2\cdot p_1}{n_1\cdot p_1}+{\cal O}(\hbar^2)\viq
d_0=-m\,\hbar \omega_2\frac{n_1\cdot n_2
}{(n_1\cdot p_1)^2}+{\cal O}(\hbar^2)\label{limd1}\\
  \bfd_\perp&=&\hbar
\,\lbrack\,\frac{ \bfk_{2\,\perp}}{n_1\cdot p_1} -\frac{n_1\cdot k_2}{(n_1\cdot p_1)^2}\,\bfp_{1\,\perp}\,\rbrack+{\cal O}(\hbar^2)=
 \frac{\hbar\,\omega_2}{mc^2}\,{\bf W}_\perp+{\cal O}(\hbar^2)\,.\label{limd2}
\eeqa
The vector $\,{\bf W}\,$ is the one defined in (\ref{W}).

With the previous results, the comparison with (\ref{fazacltr1}) shows  the expected behaviour:
\beq
\Phi(\widetilde p_2, p_1;\phi) = \Phi^{\mathrm{cl}}(p_1;\phi)+{\cal O}(\hbar)\,.
\eeq

 In the integrals $\, \bfa^{(q)}\,$ and $\,b^{(q)}\,$ defined by (\ref{ABq})  only the phase $\,\Phi\,$ is affected by the limit $\,\hbar\rar 0\,$, so one has
\beqa\label{ABcls}
\bfa^{(q)}&\rar&\bfa^{\mathrm{cl}}\equiv-\int_{-\infty}^\infty
\,d\phi\,\frac{e\bfA(\phi)}{mc}\,\exp[-i\Phi^{\mathrm{cl}}(p_1;\phi)]\,,\\
 b^{(q)}&\rar& b^{\mathrm{cl}}\equiv \int_{-\infty}^\infty
\,d\phi\,\frac{e^2\,\bfA^2(\phi)}{2m^2c^2}\,
\exp[-i\Phi^{\mathrm{cl}}(p_1;\phi)]\,.
\eeqa

Using Eq. (\ref{limd1}) in (\ref{c14}), one obtains the classical limit of the coefficients $c_1,\ldots,c_4$: 
\beq\label{coefcl}
c_1^{\mathrm{cl}}=1,\quad c_2^{\mathrm{cl}}=-1\viq c_3^{\mathrm{cl}}=c_0(2-c_0)\viq 
c_4^{\mathrm{cl}}=1-c_0\viq c_0=\frac{(mc)^2(n_1\cdot n_2)}{(n_1\cdot p_1)(n_2\cdot p_1)}.
\eeq

Using the previous expressions we get from  the quantum spectrum (\ref{quantum}), the classical limit
\begin{eqnarray}\label{class}
\frac{d^2W^{cl}}{d\Omega_2\,d\omega_2}&=& \frac{e_0^2}{4\pi^2 c}\,\omega_2^2\,\frac {(mc)^2}{(n_1\cdot p_1)^2}\{\mid \bfa^{\mathrm{cl}}\mid^2 -
\frac{(n_1\cdot p_1)^2}{(n_2\cdot p_1)^2}\mid \bfa^{\mathrm{cl}}\cdot {\bf W}\mid^2\nonumber \\&+&\,c_3^{\mathrm{cl}}\mid b^{\mathrm{cl}}\mid^2 
-2\,\frac{n_1\cdot p_1}{n_2\cdot p_1}c_4^{\mathrm{cl}}\,\Re\lbrack\left(b^{\mathrm{cl}}\right)^*\,\bfa^{\mathrm{cl}}\cdot {\bf W}\,]  \,\}\,.
\end{eqnarray}

For an electron initially at rest,  $\,\bfp_1=0\,$, one has  
$\,n_1\cdot p_1 \rar mc\,,\; n_2\cdot p_1 \rar mc \,,\; {\bf W} \rar \bfn_2\, $
and the expression of the photon distribution has the simpler form 
\beq
\frac{d^2\,W^{\mathrm{cl}}}{d\Omega_2\,d\omega_2} \rar \frac{e_0^2}{4\pi^2\,c} \,
\omega_2^2\,\left(
\mid \bfa^{\mathrm{cl}}\mid^2-
\mid \bfa^{\mathrm{cl}}\cdot\bfn_2\mid^2 +\sin^2\theta\mid b^{\mathrm{cl}}\mid^2-2\,\cos\theta\,
 \Re[\left(b^{\mathrm{cl}}\right)^*\,\bfa^{\mathrm{cl}}\cdot \bfn_2\,]\right)\,,
\eeq
with $\,\theta\,$ the photon scattering angle,  $\,\bfn_1\cdot\bfn_2=\cos\theta\,$.

Equation (\ref{class}), with the coefficients defined in  (\ref{coefcl}),  represents our final compact formula for the classical spectral and angular distribution in the case of scattering of an external  electromagnetic plane wave with arbitrary polarization on a free electron.

As mentioned in Sect. \ref{s-i}, for {\it linear polarization of the pulse, orthogonal to the electron initial momentum} the same type of formula as (\ref{class}) was established  by Krafft {\it et al} \cite{Krf1,Krf2}, by a completely different procedure.  Defining  the scattering plane as the plane containing the vectors $\,\bfp_1\,$ and $\,\bfn_1\,$, these authors start with the radiation spectrum decomposed into the contribution of polarization perpendicular ($\sigma$) and polarization parallel ($\pi$) to the plane of scattering in the reference system in which the electron is at rest;  then they  use an appropriate Lorentz transformation to describe the case of an initially moving electron. Their final results are contained in Eqs. (14-16) of \cite{Krf2}. We compared our final result (\ref{class}),  taken for  a linearly polarization pulse   with $\,\bfA\cdot\bfp_1=0$,  and the sum of the contributions of the two polarizations, as  given by the quoted authors. Agreement was found, except for the extra factor 1/2  that  Krafft {\it et al.}  use  and explain after their Eq. (1).

The derivation we presented for Eq. (\ref{class}) was  simple, but it  is based  on knowledge of  the quantum spectrum. The question arises how this formula   emerges from the general equation  (\ref{eq:S}) of CED. We  give the answer  in Appendix \ref{sa-b},  where the   expression (\ref{eq:S}), valid for an electron with an initial velocity different of zero, is transformed in the case of a pulse. Our final result,  justified in Appendix \ref{sa-b}, expresses  the vector $\,\bfK\,$   by the simpler vector $\,\bfK_0\,$, as 
\beq\label{K0K}
\bfK=\bfn_2\times(\bfn_2\times \bfK_0)\,,
\eeq
where
\beqa\label{K0}
&& \qquad  \qquad \bfK_0=-i\,\frac{e_0}{\pi\sqrt{8c}}\frac{mc}{n_1\cdot p_1}\omega_2\,\lbrack\,\bfa^{\mathrm{cl}}+t_1\,\bfn_1+t_2\,\frac{\bfp_1}{mc}\,\rbrack \,.\nonumber \\
&&t_1=\frac{mc}{n_1\cdot p_1}\left(b^{\mathrm{cl}}+\frac{\bfa^{\mathrm{cl}}
\cdot \bfp_1}{mc}\right) \viq t_2=-\frac{n_1\cdot p_1}{n_2\cdot p_{1}}\,\lbrack\,\left(\frac{mc}{n_1\cdot p_1}\right)^2n_1\cdot n_2 \,b^{\mathrm{cl}}-{\bf W}\cdot \bfa^{\mathrm{cl}}\,\rbrack\,.\qquad
\eeqa
The vector $\,{\bf W}\,$ is defined in (\ref{W}) and  the quantities $\,\bfa^{\mathrm{cl}}\,$ and $\,b^{\mathrm{cl}}\,$ are the integrals  in (\ref{ABcls}).

In this way we have derived in Appendix \ref{sa-b} an alternative expression for the vector $\,\bfK\,$ in (\ref{eq:S}), valid for the plane wave case.  From it we can calculate the emitted radiation with a given polarization. All the needed information is contained in the vector $\,\bfK_0\,$. Indeed, if $\,{\bm{\epsilon}}_1\,$  and $\,{\bm{\epsilon}}_2\,$ are two polarization vectors, 
\beq
{\bm{\epsilon}}_1^*\cdot {\bm{\epsilon}}_2=0\viq
{\bm{\epsilon}}_j\cdot \bfn_2=0\viq \mid {\bm{\epsilon}}_j\mid=1\viq j=1,2 \,,
\eeq
then one has
\beq
\bfK\cdot {\bm{\epsilon}}_j=-\bfK_0\cdot{\bm{\epsilon}}_j\viq j=1,2\,
\eeq
and 
\beq
\mid \bfK\mid^2= \mid \bfK_0\mid^2 -\mid \bfK_0\cdot \bfn_2\mid^2\,.
\eeq

The  polarization vectors chosen by Krafft {\it et  al.} are:
\begin{equation}
{\bm{\epsilon}}_{\sigma}=\frac{{\bf p}_1\times{\bf n}_2}{p_1\sin\tilde \theta},\quad {\bm{\epsilon}}_{\pi}=\frac{{\bf n}_2\times({\bf p}_1\times{\bf n}_2)}{p_1\sin\tilde \theta}\,,
\end{equation}
where $\,\tilde \theta\,$ is the angle between  the directions of the emitted photon and of the initial electron momentum.  Denoting by $ v_{\sigma}$ and $v_{\pi}$ the components of an arbitrary vector along the  polarization vectors ${\bm{\epsilon}}_\sigma$ and ${\bm{\epsilon}}_{\pi}$, respectively,   we obtain:  
\begin{eqnarray}\label{pol1}
d^2  W^{\mathrm{cl}}_{\sigma}/d\Omega_2d\omega_2&=&\frac{e_0^2}{8\pi^2c}\omega_2^2\frac{(mc)^2}{(n_1\cdot p_1)^2}\times{\vline\,- a_{\sigma}^{\mathrm{cl}}-\frac{{\bfa}^{\mathrm{cl}}\cdot{\bf p}_1}{n_1\cdot p_1} n_{1\sigma}-\frac{m\,c}{n_1\cdot p_1}\,b^{\mathrm{cl}}\, n_{1\sigma}\,\vline}^{\,2}\,,
\end{eqnarray}
\begin{eqnarray}\label{pol2}
d^2 W^{\mathrm{cl}}_\pi/d\Omega_2d\omega_2&=&\frac{e_0^2}{8\pi^2c}\omega_2^2\frac{(mc)^2}{(n_1\cdot p_1)^2}\times\\
&&\hspace*{-2cm}\times{\vline\,-a_{\pi}^{\mathrm{cl}}-\frac{n_1\cdot p_1}{n_2\cdot p_1}{\bf a}^{\mathrm{cl}}\cdot{\bf W}\,\frac{p_{1,\pi}}{mc}-\frac{{\bf a}^{\mathrm{cl}}\cdot{\bf p}_1}{n_1\cdot p_1}\,n_{1\pi}+\frac{mc}{n_1\cdot p_1}\,\,b^{\mathrm{cl}}\,[-n_{1\pi}+\frac{(n_1\cdot n_2)p_{1,\pi}}{n_2\cdot p_1}]\,\vline}^{\,2}\,.\nonumber
\end{eqnarray}

 In  the particular case of a linearly polarized pulse with the  polarization vector orthogonal to the direction of  the incident electron momentum, the two previous equations coincide with Krafft {\it et al.} results (Eqs. (14) and (15) of \cite{Krf2}), with the exception of the factor of 1/2, already mentioned. 

Our previous two equations are valid for any polarization of the laser pulse.

Polarization effects in non-linear Compton scattering of a {\it monochromatic} plane wave have been studied in detail by Ivanov {\it et al.} \cite{Ivan}   who also considered effects connected with the electron spin.

\section{Comparison of  classical and  quantum spectra. Qualitative discussion}\label{s-iv}

In order to understand the main features of the emitted radiation spectrum  and the differences between the classical and quantum results, it is useful  to consider   the limiting case of the electromagnetic monochromatic  plane wave.  In this case we use the vector potential 
\begin{equation}
{\bf A}(\phi)=A_0[\,{\bf e}_x\cos(\zeta/2)\cos(\omega_1\phi)-{\bf e}_y\sin(\zeta/2)\sin(\omega_1\phi)\,]\,,\label{A_mono}
\end{equation}
 where ${\bf e}_x$ and ${\bf e}_y$ are unit vectors  along the  principals polarization axes, orthogonal to the propagation direction vector ${\bf n}_1$, and  $\zeta$  characterizes the polarization of the laser.
 We denote by $\,k_1$ the four-vector $k_1\equiv \frac{\omega_1}{c}\,n_1\,$; in this case $\,\hbar k_1\,$  has the significance of the momentum of the photon associated with the incident wave.

  Referring  to the {\it quantum} calculations, the use of the general method for dealing with the integrals in Eqs. (\ref{integrA}) and (\ref{integrB0}) (see, for instance,  Appendix B of our previous paper \cite{BF})  leads to the  expression of the frequency and angular distribution of the radiation emitted during the time $\,\tau_0\, $, 
\begin{equation}
\frac{d^2W_{\tau_0}}{d\Omega_2 d\omega_2}=\tau_0\,\sum\limits_{N\ge1}{\cal R}^{(N)}\delta(\omega_2-\omega_2^{(N)})\label{sp_mon_q}\,,
\end{equation}
where the coefficient ${\cal R}^{(N)}$, built from generalized Bessel functions, is proportional to the transition rate of the process in which $\,N\,$ laser photons are absorbed and one photon of frequency $\omega_2^{(N)}$ is emitted. 
From the many references for this type of calculation, we quote  Lyulka \cite{Lyulka}, who  has given the explicit form of the transition rate for the general case of arbitrary initial polarization  of the electromagnetic wave. More recently,   Harvey {\it et al.} \cite{UK} and  Heinzl {\it et al.} \cite{UK-nou} have reconsidered  the particular case of circular polarization.

In the monochromatic case, the spectrum consists in an infinite series of lines whose position is given by 
\begin{equation}
\omega_2^{(N)}=N\omega_1\frac{n_1\cdot p_1}{n_2\cdot q_1+N\,n_2\cdot \hbar k_1}\,.\label{o2_mono}
\end{equation}
The four vector $\,q_1\,$ in the previous relation is the ``dressed'' electron momentum, 
\begin{equation}\label{dres}
q_1=p_1+\frac{e^2A_0^2}{4\,n_1\cdot p_1}\,n_1 \,.
\end{equation}
Since $\,q_1\,$ depends on the field intensity, if follows also that the position of the lines in the spectrum is intensity dependent.  It is important to notice that the lines are not equidistant  and that $\omega_2^{(N)}$ has a finite limit for $\,N\rightarrow\infty$,
\begin{equation}
\lim\limits_{N\rightarrow\infty}\omega_2^{(N)}=\frac{c}{\hbar}\frac{ n_1\cdot p_1}{  n_2\cdot  n_1}\equiv\omega_{{\mathrm{cut-off}}}\label{cut-off}\,,
\end{equation}
which depends on the  initial momentum  and on the initial and  final geometry.

 From Eq. (\ref{o2_mono})  it follows that in the regime $\,\hbar\omega_1\ll mc^2$,  the frequency of the emitted photons can be much larger than the laser frequency if $\, n_1\cdot p_1\,$ is large and $\,n_2\cdot q_1\,$ is small; the most favorable case  for this blue-shift of the emitted frequencies with respect to the laser frequency is that of  energetic electrons in ``head-on'' collision with the laser beam, with  the emitted photon   detected in the direction of the initial electron  momentum.

  An explicit expression for the {\it classical} spectrum in the monochromatic case was derived  by many authors, we quote only  Salamin and Faisal for  the particular cases of circular \cite{F1} and linear \cite{F3} polarization of the laser field. The classical  expression for the spectrum in the monochromatic case can be obtained by taking  directly the limit $\hbar\rightarrow 0$   of the quantum results (\ref{sp_mon_q}),  as was made  in Sect. \ref{s-iii} for a pulse. Proceeding this way, one gets
\begin{equation}
\frac{d^2W_{\tau_0}^{\mathrm{cl}}}{d\Omega_2 d\omega_2}=\tau_0\sum\limits_{N\ge1}{\cal R}^{(N)\,{\mathrm{cl}}}\delta(\omega_2-\omega_2^{(N)\,{\mathrm{cl}}})\label{sp_mon_cl}\,,
\end{equation}
where
\begin{equation}
\omega_2^{(N)\,{\mathrm{cl}}}=N\omega_1\frac{n_1\cdot p_1}{n_2\cdot q_1}\label{o2_mono-cl}
\end{equation}
and ${\cal R}^{(N)\,{\mathrm{cl}}}$ is calculated directly from its quantum counterpart ${\cal R}^{(N)}$ by taking the limit $\,\hbar\rightarrow 0$. This approach of getting the classical distributions was used by Goreslavskii {\it et al.} \cite{GRSL} for  circular polarization. These authors have also derived a closed form expression for the angular distribution, generalizing the expression given previously  for a head-on collision  \cite{Esarey}.  The expression (\ref{o2_mono-cl}) follows directly from Eq. (\ref{o2_mono}) by neglecting the term $Nn_2\cdot\hbar k_1$. This is in agreement with the way the classical spectrum follows from the quantum one, as shown in Sect. \ref{s-iii} for the case of a plane-wave pulse.

In contrast to the quantum case, in the classical limit, {\it the lines  are equidistant  and  there is no   frequency cut-off} $\,$as $\;\lim\limits_{N\rightarrow\infty}\omega_2^{(N)\,{\mathrm{cl}}}=\infty.$

 It is worth  mentioning  the  particular case   of   head-on collisions,  circular polarization of the laser  and the emitted photon  detected along the direction of the incident electron (backward scattering) when,  both in the classical \cite{Esarey} and quantum case,  the coefficients ${\cal R}^{(N)}$ vanish for $N\ge2\,$. This property can explain the behaviour  found in the laser pulse case in some of the numerical examples presented in the next section.

In the monochromatic case, the comparison between the classical and quantum results  made by  Goreslavskii {\it et al.} \cite{GRSL} has established  conditions of applicability  for the classical approximation;  with $\eta\,$ defined in (\ref{defeta}) and  $\,\gamma\equiv E_1/mc^2 $, they find the conditions: i) $\,\hbar\omega_1\gamma/mc^2\ll 1\,$ for  $\,\eta\ll 1 $,  $\,\gamma \gg \eta\,$  and  $\,\bfn_2\not=\bfn_1\,$; ii)   $\,\hbar\omega_1\eta^2\gamma/mc^2\ll 1\,$ for $\gamma\gg\eta\gg1$  and iii) $\,\hbar\omega_1\eta^3/mc^2\ll 1\,$  for $\gamma\sim\eta\gg1$.  For an electron at rest they estimate the emission angles to be of the order of $\,1/\eta\,$ and give the condition $\,\hbar\omega_1\eta^2/mc^2\ll 1\,.$  The numerical examples presented by us in the following section correspond to the regime $\gamma\gg\eta\sim1$, and electron counterpropagating with respect to the laser pulse, for which Goreslavskii's Eq. (16) reduces to $\,\hbar\omega_1\gamma/mc^2\ll 1\,$.    The latter conclusion follows also   from  the analysis of the relation between Compton and Thomson scattering made by   Heinzl {\it et al.} \cite{UK-nou}. They   have reached the conclusion that the $\,l\,$ photon absorption contribution to the Compton radiation spectrum reduces to the corresponding quantity in Thomson scattering if the parameter  $\,y_l=2\,l\,\hbar k_1\cdot p_1/ [m^2c^2(1+\eta^2)]$,  defined in their equation (18) and transcribed here in our notation,   is much smaller than 1.

For the case of a {\it finite laser pulse}, in both classical and quantum cases, the spectrum becomes continuous; the discrete lines being replaced by maxima whose shape and width depend on the shape and length of the laser pulse. The  dominant maxima are located near the corresponding positions of the lines obtained in the monochromatic case for the same initial energy of the electron and for the field intensity equal to the peak-intensity of the pulse. The  extension of the spectrum  is due to the fact that,  in contrast to the monochromatic case, during the pulse, the electron feels a varying laser intensity, between $\,0\,$ and $\,I_{max}$.  As a consequence, for a fixed observation direction,  the line corresponding to  $\omega_{ 2}^{(N)}(I_{max})$ must be replaced by a continuous distribution contained  practically between   $\omega_{ 2}^{(N)}(I_{max})$ and $\omega_{2} ^{(N)}(I=0)$.  Since $\omega_{2}^{(N)}(I)$ decreases with the increase of the intensity $I$, it follows that when the intensity of the pulse increases, the peaks in the spectrum spread toward lower frequencies than the position of the line corresponding to the null intensity limit;
one can write a simple expression of the width in terms of the difference between the  corresponding wavelengths,  
\begin{equation}
\Delta\lambda_2\equiv\lambda_2^{(N)}(I_{\mathrm{max}}) -\lambda_2^{(N)}(I=0)=\frac{\lambda_1}{N}\frac{\eta^2}4\frac{(mc)^2}{(n_1\cdot p_1)^2}\,,
\end{equation}
  with $\,\lambda_1\,$ the wavelength of the initial radiation.

In the numerical examples presented in the next section we shall compare  classical and quantum spectra obtained for a finite laser pulse. We shall consider a laser  with the  frequency in the optical domain and values of   $\eta\,$ in (\ref{defeta}) of the order of unity, which corresponds to the intensity of the order of  the atomic unit,   counterpropagating   with very  energetic incident electrons (energy in the GeV range). This case is favorable for the emission of energetic photons  and important differences between the classical and quantum case will be visible.

\section{Comparison  of  classical and  quantum spectra. Numerical results}\label{s-v}

 In this section  we compare directly  numerical results for radiation spectra in Thomson and Compton scattering. We try to connect, wherever possible,  the various features of the results to the qualitative discussion in the previous section.

   We consider a Gaussian laser pulse, described by the vector potential
\begin{equation}\label{Apot}
{\bf A}(\phi)=A_0\exp\left[-\frac{(1.1774\omega_1\phi)^2}{4\pi^2\tau^2}\right]\left[\,{\bf e}_x\cos(\omega_1\phi)\cos(\zeta/2)-{\bf e}_y\sin(\omega_1\phi)\sin(\zeta/2)\,\right].\end{equation}

All the numerical results presented in this section, except for the last one, correspond to  a circularly polarized pulse,  $\zeta=\pi/2$.
The factor $1.1774$ in the exponent of the Gaussian envelope is chosen such that the dimensionless parameter $\,\tau\,$ is the full width at half maximum $FWHM$ (measured in cycles) of the intensity profile  of the pulse. The frequency $\omega_1$  was chosen, unless otherwise stated, such that $\hbar\omega_1=1.17$ eV,  i.e., close to the fundamental frequency of the   Nd:glass laser. In all the calculations presented in this paper we choose $\,\tau=10$ cycles, which, for  the wavelength  we use corresponds  to   the FWHM of the pulse of about 35 fs.
 The peak intensity is characterized by the parameter $\,\eta$ defined in (\ref{defeta}); for $\eta=1$ and for the frequency  we have  mentioned the peak intensity of the laser pulse is $I_{\mathrm{max}}\sim6\times 10^{17}$ W/cm$^2$.

As very energetic electrons produce energetic photons, whose momentum $\,\hbar k_2\,$ is not negligible,  quantum effects should appear in the regime we consider.  In order to characterize the energy $\,E_1$ of the incident electron  we  use the dimensionless parameter  $\,\gamma={E_1}/{mc^2}\,$. With one exception, we shall consider  head-on collisions; we  present results only  for  $\gamma\ge10^3$ as  at lower electron energies, for the laser parameter considered here,  we have verified that the   classical and quantum results are practically the same. This is in agreement with the condition  of validity for classical results $\,\gamma\hbar\omega_1/(mc^2)\ll1$ given by Goreslavskii and discussed in the previous section; for  $\,\gamma=10^3$ these ratio is  0.002.

 For  the particular geometry chosen and  a circularly polarized laser pulse, the process has an axial symmetry about the laser propagation direction.  At   the very high energy considered for the incident electron, the radiation is emitted in a very narrow  cone whose axis has the  direction of the incident electron.  We shall denote by $\,\theta_{\gamma-e}=\pi-\theta\,$ the angle  between the direction of the incident electron and that of the emitted photon.

We shall start with the  illustration of a {\it scaling law} valid for the classical results,    true if the envelope in the expression of the vector potential  is chosen as a function of the product $\,\omega_1\phi\,$ only,  as it is the case in Eq. (\ref{Apot}).  In such a situation,  from Eq.  (\ref{class}) it follows,  by simply making a change of variable from $\phi\,$ to $\,\omega_1\,\phi\,$ in the integrals (\ref{ABcls})  and from  $\chi\,$ to $\,\omega_1\,\chi\,$  in the integrals in $\,\Phi^{\mathrm{cl}}(p_1;\phi)\,$, that  $\,\frac{d^2W^{\mathrm{cl}}}{d\omega_2d\Omega_2}\,$ is a function of $\,{\omega_2}/{\omega_1}\,$ only.   In order to illustrate this property,  we have calculated the spectrum  for $\gamma=10^4$,   $\,\eta=2\,$ and $\,\theta_{\gamma-e}=10^{-4}\pi$, for three values of the laser frequency $\omega_1=n\omega_0$, with $n=1$, 2, and 10 and $\hbar\omega_0=1.17$ eV. In Fig.   \ref{Figs} the spectra $\frac{d^2W}{d\Omega_2d\omega_2}$ are represented as  function of the scaled frequency $\,\omega_2/n\,$.  The  full black lines correspond to $n=1$, the dotted red to  $n=2$  and  the dashed green lines to $n=10$. The classical results should be the same in all three cases, and  as the corresponding curves overlap perfectly, we see only one curve in the upper part of the figure. The agreement between the numerical  results obtained  in the classical case for the three different incident frequencies is a check of our classical code.
In the quantum case,  since the scaling law is not valid, the three spectra are different. 
\begin{figure}
\includegraphics[scale=0.3]{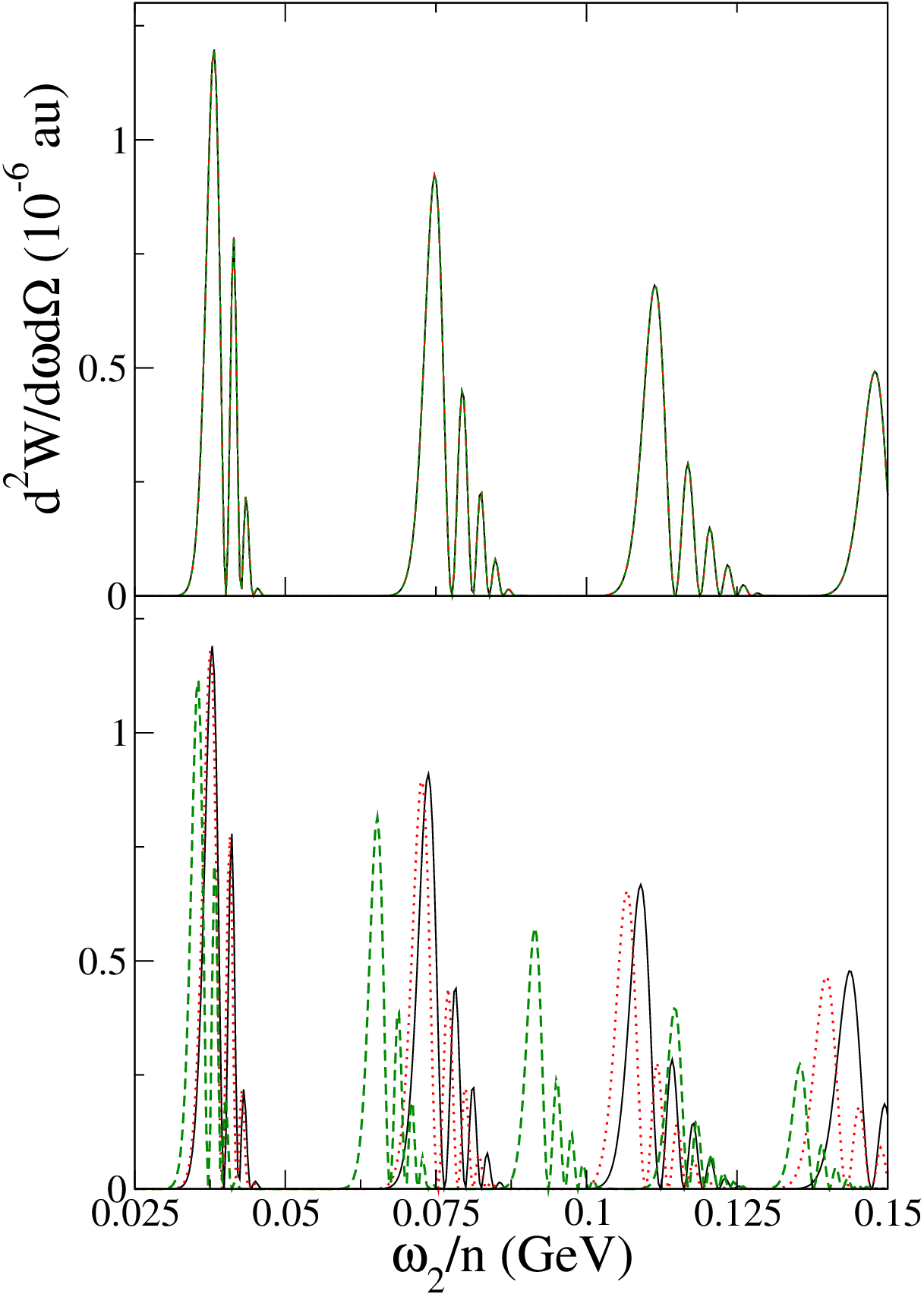}
\caption{(Color online) Double differential spectrum $\frac{d^2W}{d\Omega_2d\omega{_2}}$ as a function of $\omega_2/n$,   for incident frequencies $\,\omega_1=n \,\omega_0\,,\;\omega_0=1.17 eV$;  $n=1$ full (black) line, $n=2$ dotted (red) line, $n=10$ dashed (green) line. Upper plot: classical calculation, lower plot: quantum calculation.\label{Figs}}
\end{figure}

Next we shall  consider angular distributions $\,\frac{dW}{d\Omega_2}\,$ of the emitted photon.
In Fig. \ref{fig1}   the angular distribution as a function of the angle $\,\theta_{\gamma-e}\,$ is represented for three values of the electron initial energy: (a) $\gamma=10^3$, corresponding to the initial electron energy $E_{1}\sim511$ MeV, (b) $\gamma=10^4$ ( $E_1\sim5.1$ GeV) and  (c)  $\gamma=10^5$ ($E_1\sim51$ GeV),  and for the values 0.5, 1 and 2 of the parameter $\eta$, marked on each graph. 
\begin{figure}
\includegraphics[scale=0.45]{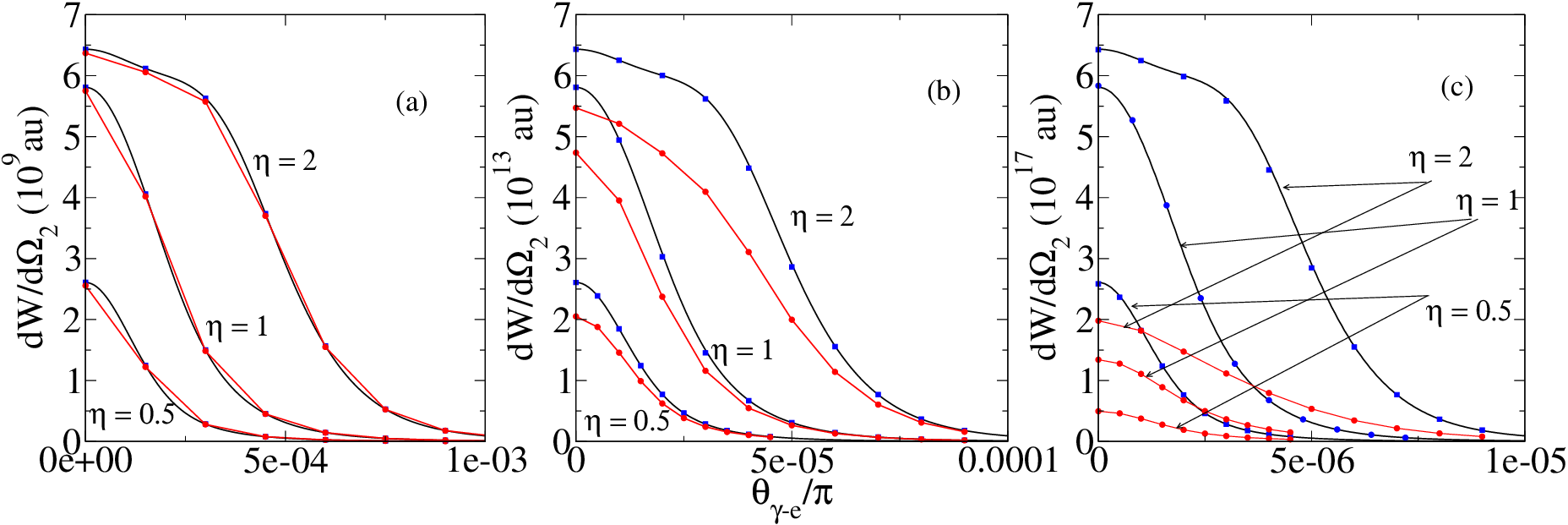}
\caption{(Color Online) The angular distribution  $\,\frac{dW}{d\Omega_2}\,$  for $\eta$=0.5, 1 and 2 (values marked on  each graph) and  $\gamma=10^3$ (a),  $\gamma=10^4$ (b),  $\gamma=10^5$  (c). Black lines: classical   results  (\ref{ang}), blue squares:  numerical integration of (\ref{class}), red circles: quantum results (\ref{quantum}).\label{fig1}} 
\end{figure}
Values calculated in the framework of the classical theory (Thomson), using the analytical formula 
(\ref{ang}), are represented with black lines.  The blue squares represent the values obtained by numerical integration  over the emitted frequency  of our formula  (\ref{class}); the agreement between the two is also a check of the accuracy of our classical code.  The quantum (Compton) results, obtained by numerical integration of the double differential  spectrum (\ref{quantum}) are represented with red circles, which have been connected by lines in order to guide the eye. For $\,\gamma=10^3$ the classical and quantum results are practically identical; differences  are found  at $\,\gamma=10^4$ and they become  large  at $\gamma=10^5$. In all cases, the classical and quantum results have similar shapes; the quantum results are always lower than the corresponding classical ones which is a consequence of the fact that in the quantum case the spectrum is compressed toward lower frequencies (see also the discussion  further on).  When the field intensity increases, the angular distribution becomes ``wider'', i.e. it spreads up to larger angles $\theta_{\gamma-e}$.  In the monochromatic case, the angular distribution  was analyzed by Goreslavskii {\it et al.} \cite{GRSL}. For $\,\gamma\gg \eta \gg 1$, they have evaluated  the opening angle  of the cone in which the radiation is emitted, as inversely proportional to the factor $\gamma_*$    associated with the ``dressed'' electron momentum.  For the  geometry chosen here,  $\gamma_*$ decreases  with increasing  $\,\eta$. 
We also note that the opening angle of the cone in which the radiation is emitted  decreases with the increase of $\gamma$  at fixed $\,\eta\,$,  in agreement  also with the  predictions made  in \cite{GRSL}.

In the following, in Figs. 3-6, we  study in  more detail the distribution of the emitted photons, by presenting  the double differential spectra $\frac{d^2W}{d\Omega_2d\omega_2}$  as a function of the emitted frequency $\omega_2$,  for a fixed value of  $\,\theta_{\gamma-e}\,$. The calculations are made for the same initial scattering geometry (head-on collision)  as in Fig. \ref{fig1}  and for the values  $10^4$ and $10^5$ of the parameter $\,\gamma$. 

 In Fig. \ref{fig2} we consider    $\,\eta=0.5$ and $\gamma=10^4$; the three plots correspond, respectively, to $\,\theta_{\gamma-e}=0$ (a), $\,\theta_{\gamma-e}=5 \times 10^{-5}\pi$ (b) and $\,\theta_{\gamma-e}=10^{-4}\pi$ (c). Classical results are drawn in full (black) lines, and the quantum ones in dashed (red) lines.  The vertical  thin lines indicate the positions of the discrete frequencies   $\omega_2^{(N)}(I_{max})$,  [Eq. (\ref{o2_mono})  for Compton  and (\ref{o2_mono-cl}) for Thomson];  in  case (a)  the values  of $\,\omega_2^{(1)}(I=0)\,$ are also shown, marked with dashed lines. For the laser pulse,  each line is replaced by a continuous distribution   with  the absolute maximum near the corresponding frequency   $\omega_2^{(N)}(I_{max})$.  If $\,\eta\,$ is not too large these  distributions   do not overlap.
\begin{figure}
\includegraphics[scale=0.5]{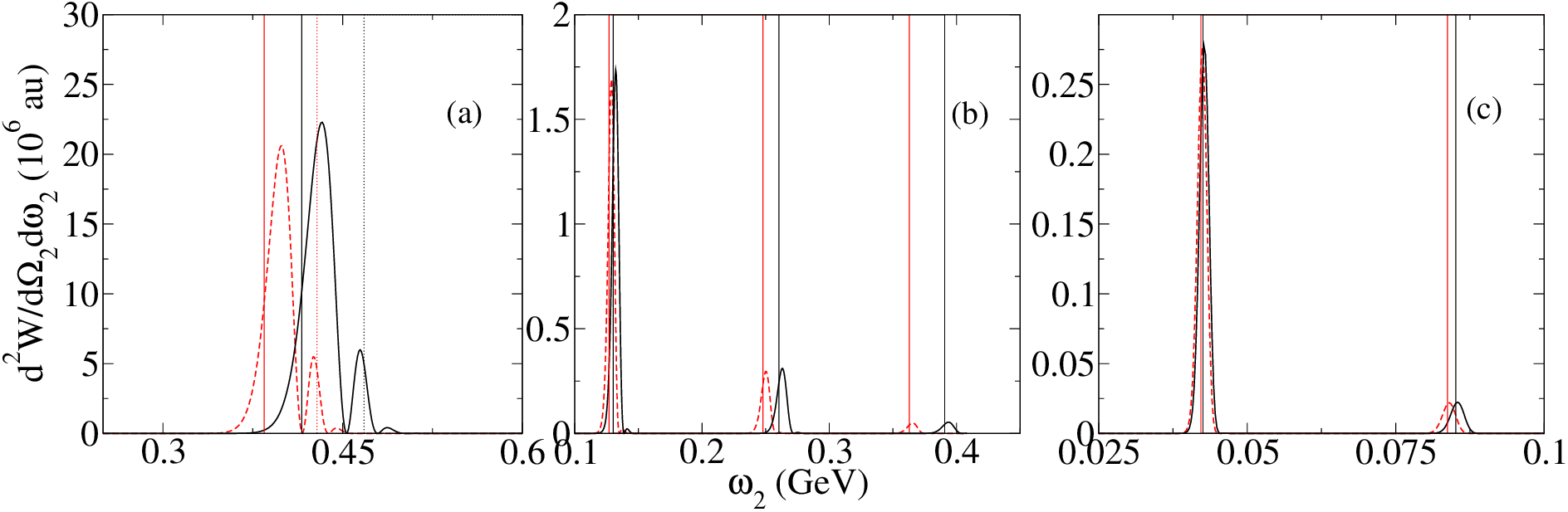}
\caption{(Color online)Double differential spectrum $\frac{d^2W}{d\Omega_2d\omega_2}$ for $\gamma=10^4$ and $\eta=0.5$. $\theta_{\gamma-e}=0$ (a), $5\times10^{-5}\pi$ (b) and $10^{-4}\pi$ (c). Full (black) line: classical calculation, dashed (red) line: quantum calculation.  Thin vertical lines: the position of $\omega_2^{(N)}(I_{\mathrm{max}})$; dashed vertical lines in (a):  the positions of $\omega_2^{(N)}(I=0)$.\label{fig2}}
\end{figure}

  We shall discuss first {\it case (a)}, the  backward emission,  because in the context  we consider (head-on collision, circularly polarized laser)  the situation differs drastically from that at  $\,\theta_{\gamma-e}\not=0\,$. In case (a) there is  only one  spectral region in the spectrum, a particularity that  has a correspondence in the monochromatic limit, where only one line ($N=1$) appears, as explained in Sect. \ref{s-iv}. The width of this maximum can be  estimated, according to the previous section as $\omega_2^{(1)}(I=0)-\omega_2^{(1)}(I_{\mathrm{max}})$; the agreement between this estimation and the numerical results is seen on the graph.

 At {\it  scattering angles $\,\theta_{\gamma-e}\not =0\,$}, both Thomson and Compton  spectra consist of a series  of successive maxima located near the corresponding monochromatic positions [different values of $N$ in (\ref{sp_mon_cl}) and (\ref{sp_mon_q})] and with a small structure around the principal maximum. Due to the relatively low value of the parameter $\,\eta$, only  a few maxima are present  in the cases (b) and (c)  of  Fig. \ref{fig2}, and  the first one is dominant. We notice that the new spectral regions appear as soon as $\,\theta_{\gamma-e}\,$ departs from 0, so they could influence the experimental observation of backward scattering at high intensity. 

    More maxima appear, except for the case (a), for a higher field intensity than in Fig. \ref{fig2}, as shown in  Fig.  \ref{fig4},  corresponding to $\,\eta=2$;
\begin{figure}
\includegraphics[scale=0.5]{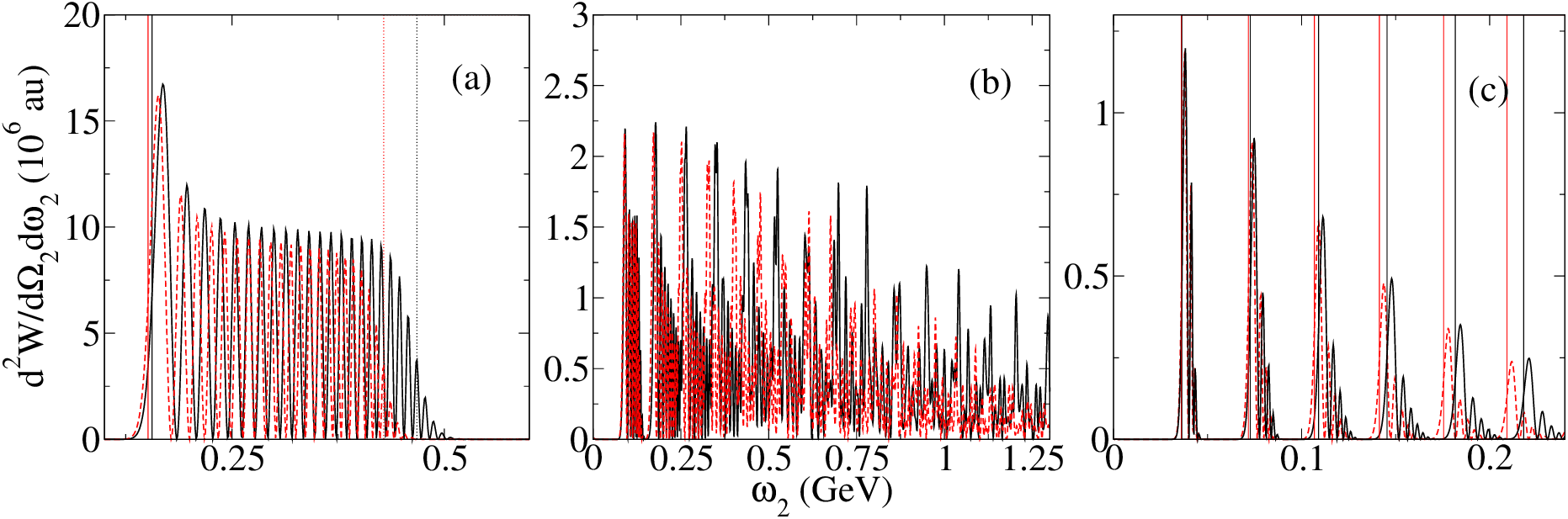}
\caption{Same as \ref{fig2} but $\eta=2$\label{fig4}}
\end{figure}
  also their structure becomes wider.  The spectral region covered at $\theta_{\gamma-e}=0$  becomes  very wide;   as in Fig. \ref{fig2}, its width can be explained by the variable intensity during the pulse.  For  $\theta_{\gamma-e}=5 \times 10^{-5}\pi\,$ the width of the successive maxima is so large that they practically overlap.

For increased electron energy ($\gamma=10^5$) more important differences between the classical and quantum results appear, in agreement  with  the observations made for the angular distribution $\frac{dW}{d\Omega_2}$,  in connection with Fig. \ref{fig1}. Figures \ref{fig6}  and \ref{fig8} correspond to the same values for $\eta\,$ as Fig. \ref{fig2} and, respectively, Fig. \ref{fig4}, but  $\gamma$ is $10^5$ and the angle $\,\theta_{\gamma-e}\,$ ten times smaller.
The qualitative behaviour of the results is the same,  nevertheless,  on the one hand   the shift of the quantum results with respect to the classical ones is much larger  and on the other hand, the classical spectrum spreads to much larger frequencies than the quantum one,  which explains the smaller value of the frequency integrated spectrum ($\,dW/d\Omega_2$, Fig. \ref{fig1}) for Compton than for Thomson scattering. 
\begin{figure}
\includegraphics[scale=0.5]{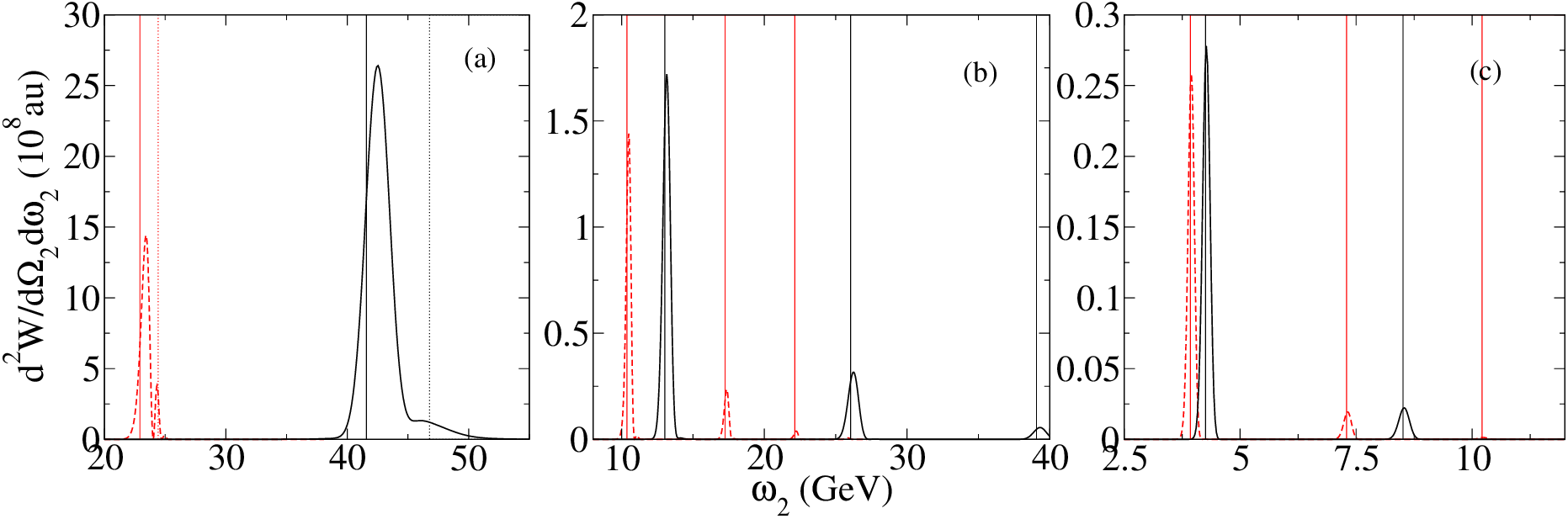}
\caption{(Color online) Double differential spectrum $\frac{d^2W}{d\Omega_2d\omega_2}$ for $\gamma=10^5$ and $\eta=0.5$.  $\theta_{\gamma-e}=0$ (a), $5\times10^{-6}\pi$ (b) and $10^{-5}\pi$ (c). Dotted (red) line: classical calculation, full (black) line: quantum calculation.  Thin vertical lines mark the position of $\omega_2^{(N)}(I_{\mathrm{max}})$; dashed vertical lines in (a) mark the positions of $\omega_2^{(N)}(I=0)$}\label{fig6}
\end{figure}
\begin{figure}
\includegraphics[scale=0.5]{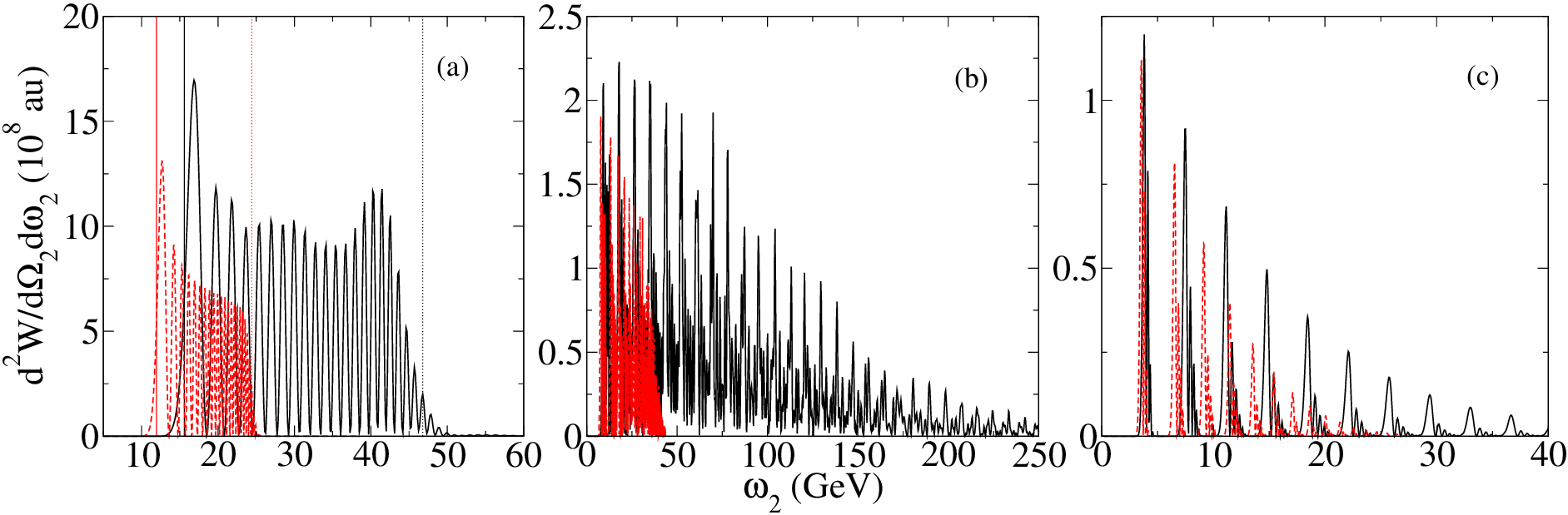}
\caption{Same as \ref{fig6} but $\eta=2$\label{fig8}}
\end{figure}

 For  the larger intensity $\eta=2$  in  Fig. \ref{fig8},  the    position of the maxima  is very different in classical and quantum cases, as it is for the line positions  in the monochromatic case, and  the width of the peaks  becomes very wide;  the lower and upper limits for frequencies are close to  the values for the maximum and zero intensity in the monochromatic case, marked with vertical lines on the graph (a)  only.  With increasing $\,\theta_{\gamma-e}\,$ the spectral region covered by radiation emission becomes narrower  and it moves toward lower frequencies. 

 As we have just mentioned, the most important differences between classical and quantum  predictions, are   visible in Fig. \ref{fig8}.  Although at first  sight, the case (a) seems similar to the other cases, it is not so.  As already explained, all that is seen in case (a) corresponds to a line in the monochromatic case, that with $\,N=1\,$, in both Thomson and Compton scattering. An increase of $\eta\,$ will result in a broadening of the spectrum toward low frequencies, as according to (\ref{sp_mon_q}) $\omega_2\,$ decreases with the intensity. In the  case (b)   both in Compton and Thomson case many overlapping maxima are present; also the Compton  results indicate   an upper} limit  of the frequency for the whole spectrum, corresponding in the monochromatic case to  the cut-off frequency  (\ref{cut-off}), independent on intensity.  For $\,\gamma=10^5$ this limit is  $\omega_{\mathrm{cut-off}}\sim51$ GeV, in both cases, as the  two angles are very close. This value is practically equal to the kinetic energy of the incident electron. A further  increase of the intensity  will lead only to an increase of the intensity of this portion of the  emitted spectrum, but will not extend it to  larger frequencies. The value of the limit frequency can be changed only through the incident electron momentum magnitude and direction.
For $\theta_{\gamma-e}=5\times10^{-6}$ the classical spectrum spreads up to $\omega_2\sim150$ GeV, while for $\theta_{\gamma-e}=5\times10^{-6}$ the limit is $\omega_2\sim250$ GeV. This limit is due only to the finite value of the parameter $\eta$;  Were  $\,\eta\,$ larger, additional maxima, located at larger values of $\omega_2$ would appear, as there is no limit to the frequency spectrum in Thomson scattering.  One can conclude that in the studied conditions, the classical results cease to be valid.

Finally, we present an example of polarization analysis of the emitted spectrum for  Thomson scattering.  We now choose a geometry different from  that used in the previous cases,  with the  incident electron momentum   along the $\,Ox$ axis and orthogonal to the  direction of propagation of the laser  
(always the $\,Oz$ axis). We have chosen $\,\gamma=10^3$, $\,\eta=1$  and the same frequency $\,\omega_1\,$, i.e. $\,\hbar \omega_1=1.17$ eV,  as in the previous cases. The emitted photon is detected in the plane defined by  the direction of propagation of the laser and  the  direction of the incident electron, at an angle $\theta_{e-\gamma}=2.5\times10^{-4}\pi$ with respect to  ${\bf p}_1$. In Fig. \ref{fig-pol} are presented the spectral distributions of the two components of the emitted radiations $\,d^2W_{\sigma}^{\mathrm{cl}}/d\omega_2d\Omega_2$ (dashed green line) and $\,d^2W_{\pi}^{\mathrm{cl}}/d\omega_2d\Omega_2$ (full blue line) defined as in Eqs. (\ref{pol1}) and (\ref{pol2}), respectively.  The results change in a visible manner with  the laser polarization.  In the case (a), when the laser is polarized along the electron initial direction,  which corresponds to the choice $\,\zeta=0$ in Eq. (\ref{Apot}),   only the  $\,\pi\,$   component is present in the spectrum.  If the laser is circularly polarized  [$\zeta=\pi/2$, case (b)], both components are present,  the $\sigma$ component  being favored. Finally, the {graph}  (c) is for the laser field linearly polarized on the direction orthogonal to the initial electron direction ( $\zeta=\pi$),  the geometry studied   by Krafft {\it et al.} \cite{Krf2}. In this case the polarization of the successive peaks alternates:   those with odd order  have the $\,\pi$ polarization, while the even order peaks are $\,\sigma$ polarized.  
\begin{figure}
\includegraphics[scale=0.5]{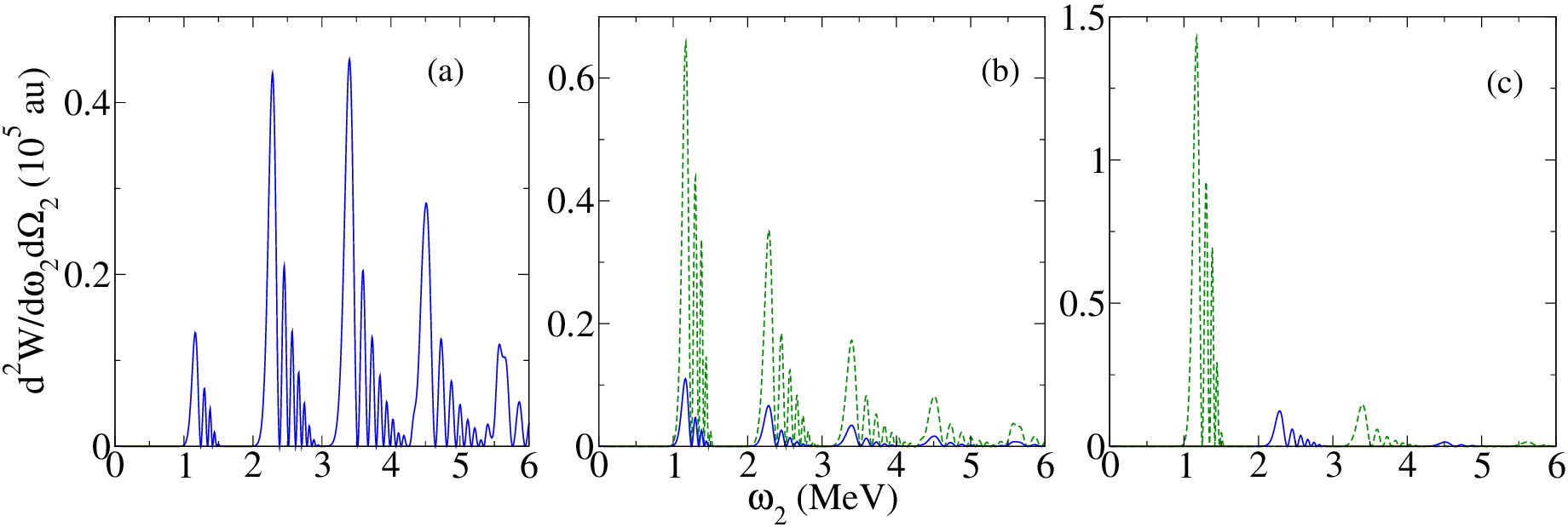}
\caption{(Color online) $d^2W_{\sigma}^{\mathrm{cl}}/d\omega_2d\Omega_2$ [dashed (green) line] and   $d^2W_{\pi}^{\mathrm{cl}}/d\omega_2d\Omega_2$ [full (blue) line] for $\gamma=10^3\,,$  $\eta=1$ and  different laser polarizations:  $\zeta=0$ (a),  $\zeta=\pi/2$ (b),  $\zeta=\pi\,$ (c)  in Eq. (\ref{Apot}).\label{fig-pol}}
\end{figure}

\section{Conclusions}\label{s-vi}

We have considered the basic equations of Compton and Thomson scattering of a non-monochromatic electromagnetic plane wave on free electrons.    We have derived the classical limit  (\ref{class}) of the quantum expression of the emitted radiation spectral and angular distribution  (\ref{quantum}), valid for any  polarization of the  laser,  for unpolarized initial electrons and summed over scattered electron spin and emitted photon polarization.  In Appendix \ref{sa-b},  we have reobtained  the expression (\ref{class}) by transforming the standard more general formula of classical electrodynamics (\ref{eq:S}). This latter calculation also  allows  the derivation of the contribution to the total distribution of the  radiation with  fixed polarization to be achieved.  Analytic expressions for two particular polarization contributions are given in (\ref{pol1}) and (\ref{pol2}).   These results are of the same type as those obtained by Krafft {\it et al.} \cite{Krf2} with a different analytical procedure and valid only  for a linearly polarized pulse orthogonal to the initial electron direction. The differences between quantum and classical spectra, qualitatively discussed  in Sect. \ref{s-iv}, are illustrated by several graphs in Sect. \ref{s-v} for the case of a  circularly polarized pulse,  head-on collisions and very energetic incident electrons. For electron energies above 500 MeV quantitative differences between classical and quantum results were shown.  The influence of the laser polarization on the emitted radiation polarization is illustrated in Fig. \ref{fig-pol}, for the case of  Thomson scattering.

\acknowledgments

This work was supported by CNCSIS-UEFISCSU, project number 488 PNII-IDEI 1909/2008. 
The authors acknowledge useful discussions with Mihai  Dondera and Victor Dinu. This work started and the bibliography was updated during a short visit of one of the authors (V.F.) at the Institute for Theoretical Atomic, Molecular and Optical Physics in Cambridge, Ma., supported by the visitor program of this Institute.  

\appendix

 \section{The electron trajectory}\label{sa-a}

In a system of axis with the origin located at the initial position of the electron and the $Oz$ axis along the propagation direction of the laser pulse (the unity vector $\,\bfn_1\,$), the electron trajectory follows from the well-known equations \cite{F1} 
\beqa\label{traj} 
z(t)&=&\frac{c}{2(n_1\cdot p_1)^2}\,\int_{\phi_0}^{\phi} d\chi\,\left\{\lbrack\,\bfp_{1\,\perp}-e\bfA(\chi)\,\rbrack^2+(mc)^2-(n_1\cdot p_1)^2\,\right\} \nonumber \\
 \bfr_\perp(t)&=&\frac{c}{n_1\cdot p_1}\,\int_{\phi_0}^\phi d\chi\,\lbrack\,\bfp_{1\,\perp}-e\bfA(\chi)\,\rbrack\viq \phi=t-\frac{z}{c}\,.
\eeqa
The first relation is an implicit equation from which one derives $\,z(t)\,$. Once this   equation is solved, the position $\,\bfr_\perp\,$  in the plane orthogonal to the direction of propagation follows.
 As for  $\,\phi=\phi_0\,$ one has $\,\bfr=0\,$, it follows that the value of $\,\phi_0\,$ coincides with the initial moment  at which the electron is supposed to be free.

 From (\ref{traj}) one obtains  the velocity:
\beq\label{eq:beta}
\bfbe=\frac{\bfv}{c}=\,\frac{1}{c}\,\frac{d\bfr}{d\phi}\frac{d\phi}{dt}=\{\frac{{
\bf p}_{1\perp}-e{\bf A}(\phi)}{n_1\cdot p_1}+\frac{\lbrack\,{\bf p}_{1\perp}-e{\bf A}(\chi)\,\rbrack^2-(n_1\cdot p_1)^2+(mc)^2}{2(n_1\cdot p_1)^2}\,\bfn_1\,\}\, \frac{d\phi}{dt} \\
\eeq
with
\beq\label{dtf}
 \frac{d\phi}{dt}=
\frac{2(n_1\cdot p_1)^2}{\lbrack\,{\bf p}_{1\perp}-e{\bf A}(\chi)\,\rbrack^2+(n_1\cdot p_1)^2+(mc)^2}\equiv F\,.\\
\eeq
For $\,\bfA=0\,$ the velocity reduces to $\,c^2\,\bfp_1/E_1\,,$ with $\,E_1=c\sqrt{m^2c^2+\bfp_1^2}\,$. So, in the case of a laser pulse, the velocity at the end of the pulse is the same as the beginning of the pulse,  
\beq\label{bif}
\bfbe(t_f)=\bfbe(t_{\mathrm{in}})=c\frac{\bfp_1}{E_1}\,.
\eeq

But, as is also known, the  equations (\ref{traj}) allow the coordinates to be treated as depending explicitly on the variable  $\,\phi\,$.  These relations have been used in transforming the classical phase from (\ref{fazacl}) to (\ref {fazacltr1}).

\section{ Transformation of Eq. (\ref{eq:S})}\label{sa-b}

We transform here the expression (\ref{eq:S}) of the classical spectrum. First, in order to avoid possible divergent integrals in the intermediate calculation,  we define the integral $\,\tilde \bfK\,$  connected to the vector $\,\bfK\,$ by
\beq
\bfK=\lim_{t_1\rar -\infty\; t_2\rar\infty} \left(\frac{e_0}{\pi\,\sqrt{8c}}\,
 \widetilde {\bf K}(t_1,t_2;\omega_2,\bfn_2)\right)\,,
\eeq
where
\begin{equation}\label{smic}
\widetilde   {\bf K}(t_1,t_2;\omega_2
,\bfn_2)  =\int_{t_1}
^{t_2}
\,\frac{dt}{\kappa^2}\,\exp[i\omega_2(t-\frac{{\bf n}_2\cdot{\bf r}(t)}{c})]\,{\bf n}_2\times[({\bf n}_2-\bfbe)\times\dot{\bfbe}]\,.
\end{equation}
 We assume that the electromagnetic field is zero outside the interval $\,(t_1,t_2)\,$.

We perform several transformations. 

1) The standard {\it integration by parts} \cite{Jks}, based on
$$\frac1{\kappa^2}{\bf n}_2\times[({\bf n}_2-\bfbe)\times\dot{\bfbe}]=\frac{d}{dt}\frac{\bfn_2\times(\bfn_2\times\bfbe)}{\kappa}\,,
$$
leads to the structure 
\beqa
\widetilde {\bf K}&=&{\bf n}_2\times({\bf n}_2 \times {\bf L}) \viq {\bf L}={\bf L}_1+
{\bf L}_2\, \\ {\bf L}_1&=&{\frac{\bfbe}{\kappa}\,\exp[i\omega_2(t-\frac{{\bf n}_2\cdot{\bf r}(t)}{c})]\vline}_{\;t_1}^{\;t_2}\\
  {\bf L}_2&=&-{ i}\,\omega_2\int_{t_1}
^{t_2}
\,dt\,\bfbe\,\exp[i\omega_2(t-\frac{{\bf n}_2\cdot{\bf r}(t)}{c})]\, \,.
\eeqa
Both vectors $\,{\bf L}_1\,$ and  $\,{\bf L}_2\,$ are in fact not well defined if the integration limits $\,t_1\,$ and $\,t_2\,$ go to $\,-\infty\,$ and to $\,
\infty\,$, respectively.  Nevertheless, for the electron initially at rest, as the velocity will be zero at the end of pulse, too, one has $\,{\bf L}_1=0\,$ and one gets the formula (\ref{eq:S-red}).

2) In the term $\,{\bf L}_2\,$ we make {\it the change from the variable $\,t\,$ to  the
variable $\,\phi=t-\bfr\cdot\bfn_1/c\,$}. We use  the expression of  $\,d\phi/dt
\,$ in (\ref{dtf}) and we write  the phase $\,\Phi^{\mathrm{cl}}(p_1;\phi)\,$  as in (\ref{fazacltr1}). The integration limits become $\,\phi_1=t_1-z(t_1)/c\,$ and $\,\phi_2=t_2-z(t_2)/c\,$.

3) Using (\ref{eq:beta}), we write the expression of $\,\bfbe\,$ as 
$$ \bfbe= F\left(\frac{\bfp_{1\perp}-e\bfA}{n_1\cdot p_1} +\frac{1-F}{F}\,\bfn_1\right)\,,$$
 then   {\it  we split $\,\bfbe\,$ as}
\beq
\frac{\bfbe}{F}=\frac{\bfp_1}{n_1\cdot p_1}+\frac{\widetilde{\bfbe}}{F}\viq \frac{\widetilde{\bfbe}}{F}=-\frac{e\bfA}{n_1\cdot p_1}+ \left(-2\,e\bfA\cdot\bfp_1+e^2\bfA^2\right)\,
\frac{\bfn_1}{2(n_1\cdot p_1)^2}\,,
\eeq
isolating the term that does not vanish in the absence of the electromagnetic field.  Accordingly to this, we also  split the integral $\,{\bf L}_2\,$ as
$${\bf L}_2={\bf L}_{2a}+{\bf L }_{2b}\,$$
with
\beq
{\bf L}_{2a}=-i\,\omega_2\,\frac{\bfp_1}{n_1\cdot p_1}\,\int_{\phi_1}^{\phi_2}d\phi\,
\exp[-i\Phi^{\mathrm{cl}}(p_1;\phi)]\viq
{\bf L}_{2b}= -i\,\omega_2\,\int_{\phi_1}^{\phi_2}\,d\phi\,
\frac{\widetilde{\bfbe}}{F}\,\exp[-i\Phi^{\mathrm{cl}}(p_1;\phi)]\,.
\label{s2b}
\eeq

4) We perform {\it an integration by  parts} in $\,{\bf L}_{2a}\,$  based on 
$$ \exp\left(i\omega_2\phi\frac{n_2\cdot p_1}{n_1\cdot p_1}\right)= -\frac{ i}{\omega_2}\,\frac{n_1\cdot p_1}{n_2\cdot p_1}\,\frac{d}{d\phi}\exp\left(i\omega_2\phi\frac{n_2\cdot p_1}{n_1\cdot p_1}\right)\,, $$ leading to
$${\bf L}_{2a}= {\bf L}_{2a}^{(1)}+{\bf L}_{2a}^{(2)}\,,$$
with 
$${\bf L}_{2a}^{(1)}=-{\frac{\bfp_1}{n_2\cdot p_1}\, \, \exp[-i\,\Phi^{\mathrm{cl}}(p_1;\phi)]\vline}_{\;\phi_1}^{\;\phi_2}$$
and 
\beq\label{s2a2}
{\bf L}_{2a}^{(2)}=i\,\frac{\omega_2\,\bfp_1}{2n_1\cdot p_1}\,\int_{\phi_1}^{\phi_2} d\phi\,\,\lbrack\,\frac{n_1\cdot n_2}{2(n_1\cdot p_1)^2}\,e^2\bfA^2+e{\bf W}\cdot \bfA\,\rbrack\,\exp[-i\,\Phi^{\mathrm{cl}}(p_1;\phi)] \,,
\eeq
 with $\,{\bf W}\,$ defined in (\ref{W}).

{\bf 5)} One notices easily that   $\quad {\bf L}_1+{\bf L}_{2a}^{(1)}=0\,$ , as  the velocity of the electron is the same at the beginning and at the end of the pulse [see (\ref{bif})]. So the vector $\,\bf L\,$ in (\ref{smic}) we are looking for is expressed as 
\beq 
\widetilde \bfK(t_1,t_2;\omega_2,\bfn_2)= \bfn_2\times\lbrack\bfn _2\times({\bf L}_{2a}^{(2)}+{\bf L}_{2b})\rbrack\,,
\eeq
where $\,{\bf L}_{2a}^{(2)}\,$ and $\,{\bf L}_{2b}\,$  given by (\ref{s2a2}) and (\ref{s2b}), respectively.

 In the  expression we have derived  for  $\,\widetilde \bfK (t_1,t_2;\omega_2,\bfn_2)\,$ we can take now the limits   $\,\phi_1\rar -\infty\,$ and $\,\phi_2 \rar \infty\,$, corresponding to  $\,t_1\rar -\infty\,$ and $\,t_2 \rar \infty\,$. By the transformations performed we have justified the final result, Eqs.  (\ref{K0K}) and (\ref{K0}), for the vector $\,\bfK\,$ in (\ref{eq:S}),   valid for an electromagnetic  plane wave.


\begin{thebibliography} {14}
\bibitem{Eberly} J. H. Eberly, {\it Progress in Optics}, edited by E. Wolf, North Holland, Amsterdam 1969. 
\bibitem{SLAC} C. Bula {\it et al.}, Phys. Rev. Lett {\bf 76}, 3116 (1996); 
 C. Bamber {\it et al.}, Phys. Rev. D {\bf 60}, 092004 (1999).
\bibitem{Hatz} Y. I. Salamin, S. X. Hu, K. Z. Hatsagortsyan and C. H. Keitel, Physics Reports {\bf 427}, 41 (2006).
\bibitem{EKK} F. Ehlotzky, K. Krajewska and J. Z. Kaminski, {\it Rep. Prog. Phys.} {\bf 72}, 046401 (2009).
\bibitem{Esarey} E. Esarey, S. K. Ride, Ph. Sprangle, Phys. Rev. E {\bf 48}, 3003 (1993).
\bibitem{as}K. Lee, Y. H. Cha, M. S. Shin, B. H. Kim, and D. Kim, Phys. Rev. E {\bf 67}, 026502 (2003).
\bibitem{LLCW}P.F.  Lan, P.X. Lu, W. Cao, and X.L. Wang, Phys. Rev. E {\bf 72}, 066501 (2005).
\bibitem{Bab} M. Babzien {\it et al.}, Phys. Rev. Lett. {\bf 96}, 054802 (2006).
\bibitem{MKM} C. I. Moore, J. P. Knauer, and D. D. Meyerhofer, Phys. Rev. Lett. {\bf 74}, 2439 (1995).
\bibitem{Jks}  J. D. Jackson,  {\it  Classical Electrodynamics}, third edition, Wiley, 1998.
\bibitem{Krf1} G. A. Krafft, Phys. Rev. Lett. {\bf 92}, 204802 (2004).
\bibitem{Krf2}  G. A. Krafft, A. Doyuran and J. B. Rosenzweig,  Phys. Rev. E {\bf 72}, 056502 (2005). 
\bibitem{Hart3D} F. V. Hartemann, J. R. Van Meter, A. L. Troha, E. C. Landahl, N. C. Luhmann, H. A.  Baldis, A. Gupta, A. K. Kerman, Phys. Rev. E {\bf 58}, 5001 (1998).
\bibitem{cor}  K. Lee and Y. H. Cha, M. S. Shin, B. H. Kim, and D. Kim,  Optics Express  {\bf 11}, 309 (2003).
\bibitem{Gao} J. Gao, J. Phys. B: At. Mol. Opt. Phys. {\bf 39}, 1345 (2006).
\bibitem{Lan} P.F. Lan, P.X. Lu, W. Cao, and X.L. Wang, J.Phys. B: At. Mol. Opt. Phys. {\bf 40}, 403 (2007).
\bibitem{UK-nou} T. Heinzl, D. Seipt, and B. K¨ampfer, Phys. Rev. A {\bf 81}, 022125 (2010).
\bibitem{Hartpro} F. V. Hartemann, H. A. Baldis, A. K. Kerman, A. Le Foll, N. C. Luhmann, Jr., and B. Rupp, Phys. Rev. E {\bf 64}, 016501 (2001).
\bibitem{HG}F. V. Hartemann, D. J. Gibson and 
A. K. Kerman, Phys. Rev. E {\bf 72}, 026502 (2005).
\bibitem{UK} C. Harvey, T. Heinzl and A. Ilderton,  Phys. Rev. A {\bf 79}, 063407 (2009).
\bibitem {HI} T. Heinzl and A. Ilderton, Eur. Phys. J. D. {\bf 55}, 359 (2009).
\bibitem{Fofa} N. B. Narozhnyi and M. S. Fofanov, JETP {\bf 83}, 14 (1996).
\bibitem{BF} M. Boca and V. Florescu, Phys. Rev. A {\bf 80}, 053403 (2009);
 M. Boca and V. Florescu, Phys. Rev. A {\bf 81}, 039901(E) (2010).
\bibitem{Krekora} P. Krekora, R. E. Wagner, Q. Su, and R. Grobe, Laser Phys. {\bf 12}, 455 (2002).
\bibitem{wpk} J. Peatross, C. Mu¨ller, K. Z. Hatsagortsyan, and C. H. Keitel, Phys. Rev. Lett. {\bf 100}, 153601 (2008).
\bibitem{Kramers} H. A. Kramers, Phil. Mag. {\bf 46}, 836 (1923).
\bibitem{RHP} L. Kim and R. H. Pratt, Phys. Rev. A {\bf 36}, 45 (1987). 
\bibitem{GRSL} S. P. Goreslavski, S. V. Popruzhenko and O. V. Shcherbachev, Laser Phys. {\bf 9}, 1039 (1999).
\bibitem{Hart} V. Hartemann, Nucl. Instr. and Methods in Physics Research, A {\bf 608}, S1 (2009).
\bibitem{Heinzl} T. Heinzl, Journal of Physics: Conference Series {\bf 198}, 012005 (2009).
\bibitem{Sokol} I. V. Sokolov, J. A. Nees, V. P. Yanovsky, N. M. Naumova and G. A. Mourou, Phys. Rev. E {\bf 81}, 036412 (2010).
\bibitem{BK} L. S. Brown and T. W. B. Kibble, Phys. Rev. {\bf 133}, A705 (1964).
\bibitem{Sa-Sa} E. S. Sarachik and G. T. Schappert, Phys. Rev. D {\bf 1}, 2738 (1970).
\bibitem{Ivan} D. Yu. Ivanov, G. L. Kotkin and V. G. Serbo, Eur. Phys. J. C {\bf 36}, 127 (2004).
\bibitem{Lyulka}  V. A. Lyulka, Zh. Eksp. Teor. Fiz. {\bf 67}, 1638 (1974) [Sov. Phys. JETP {\bf 40}, 815 (1975)]. 
\bibitem{F1} Y. I. Salamin and F. H. M. Faisal, Phys. Rev. A {\bf 54}, 4383 (1996); ibidem  Phys. Rev. A {\bf 55}, 3964 (1997).
\bibitem{F3} Y. I. Salamin and F. H. M. Faisal, J. Phys. A: Math. Gen. {\bf 31}, 1319 (1998).
\end{thebibliography}
\end{document}